\documentclass[aps,rmp,twocolumn,groupedaddress,floatfix,notitlepage]{revtex4-1}
\pdfoutput=1
\usepackage{natbib}
\setcitestyle{authoryear}
\usepackage{amsmath}
\usepackage{amsfonts}
\usepackage{amssymb}
\usepackage{graphicx}

\newcommand{\eq}[1]{Eq.~(\ref{#1})}
\newcommand{\Eq}[1]{Equation~(\ref{#1})}
\newcommand{\fig}[1]{Fig.~\ref{#1}}

\newcommand{\eqs}[2]{Eqs.~(\ref{#1}) and (\ref{#2})}

\newcommand{\consecutiveeqs}[2]{Eqs.~(\ref{#1}-\ref{#2})}

\newcommand{\app}[1]{Appendix~\ref{#1}}

\def \pfix{p_\mathrm{fix}}

\def \Ub{U_b}
\def \Ud{U_d}
\def \Ui{U_i}
\def \sd{s_d}
\def \sb{s_b}
\def \sdc{\sd^*}
\def \sdb{\sd^\dagger}
\def \sdbar{\overline{s}_d}
\def \sbbar{\overline{s}_b}

\def \rhob{\rho_b}
\def \rhod{\rho_d}
\def \rhoi{\rho_i}

\def \xc{x_c}

\def \vo{v_0}
\def \xco{x_{c}}

\def \fo{f_0}
\def \wo{w_0}
\def \w1{w_1}

\def \F{F}
\def \xmin{x_\mathrm{min}}

\def \Rb{R_b}
\def \Rd{R_d}
\def \Rbo{\Rb^0}

\def \etarb{\Delta r_b}
\def \etard{\Delta r_d}
 
\def \xco{{x_c^0}}

\def \tsel{T_d}
\def \tc{T_c}
\def \tb{T_b}
\def \td{T_d}
\def \tsw{T_\mathrm{sweep}}
\def \tfix{T_\mathrm{fix}} 
\def \twait{T_\mathrm{wait}}
\def \test{T_\mathrm{est}}

\def \geta{{g_\eta}}
\def \gs{{g_s}}
\def \heta{{h_\eta}}
\def \hs{{h_s}}

\begin{document}

\title{Deleterious passengers in adapting populations}
\author{Benjamin H. Good$^{1}$}
\author{Michael M. Desai$^{1}$}
\affiliation{\mbox{${}^1$Department of Organismic and Evolutionary Biology, Department of Physics, and} \mbox{FAS Center for Systems Biology, Harvard University}}

\begin{abstract}

Most new mutations are deleterious and are eventually eliminated by natural selection. But in an adapting population, the rapid amplification of beneficial mutations can hinder the removal of deleterious variants in nearby regions of the genome, altering the patterns of sequence evolution. Here, we analyze the interactions between beneficial ``driver'' mutations and linked deleterious ``passengers'' during the course of adaptation. We derive analytical expressions for the substitution rate of a deleterious mutation as a function of its fitness cost, as well as the reduction in the beneficial substitution rate due to the genetic load of the passengers. We find that the fate of each deleterious mutation varies dramatically with the rate and spectrum of beneficial mutations, with a non-monotonic dependence on both the population size and the rate of adaptation. By quantifying this dependence, our results allow us to estimate which deleterious mutations will be likely to fix, and how many of these mutations must arise before the progress of adaptation is significantly reduced. 
\end{abstract}

\maketitle

\section*{Introduction}

\noindent Recent years have witnessed an increased interest in the evolutionary dynamics of rapid adaptation. Once regarded as an obscure limit of population genetics, this regime has since been observed in a variety of empirical settings, from laboratory evolution experiments~\citep{barrick:etal:2009, lang:etal:2013} to natural populations of pathogenic viruses~\citep{strelkowa:laessig:2012}, bacteria~\citep{lieberman:etal:2014}, and certain cancers~\citep{nik-zainal:etal:2012}. In these populations, natural selection plays a central role in driving beneficial variants to fixation, and significant theoretical and empirical effort has been devoted to the study of these beneficial mutations and the dynamics by which they spread through the population [see \citet{sniegowski:gerrish:2010} for a review]. Yet even in the most rapidly adapting populations, the vast majority of new mutations are neutral or deleterious, and much less is known about how these variants influence (or are influenced by) adaptation in nearby regions of the genome. As a result, even the most basic questions about this process remain unanswered. How deleterious must a mutation be before it is effectively purged by selection? Which deleterious mutations have the largest influence on the spread of the adaptive mutations? And how do the answers to these questions depend on the size of the population and the spectrum of beneficial mutations? These questions are the focus of the present study.

In the absence of other mutations, the fate of a deleterious variant is determined by the interplay between natural selection and genetic drift. Selection purges harmful variants from the population on a timescale inversely proportional to the fitness cost, $\sd$, of the deleterious mutation. Meanwhile, random fluctuations from genetic drift can drive these variants to fixation, which requires a time proportional to the effective population size, $N_e$. Deleterious mutations with $\sd \gg N_e^{-1}$ will be purged long before they can fluctuate to high frequency, while mutations with $\sd \ll N_e^{-1}$ will barely feel the effects of selection before they fix. The presence of this ``drift-barrier'' at $\sdc \sim N_e^{-1}$ has long been recognized \citep{kimura:1968, king:jukes:1969, ohta:1973}. It suggests that fewer deleterious mutations will accumulate in larger populations, and that those that do fix will have a smaller effect on fitness. However, even a small number of beneficial mutations can change this picture considerably. 

When beneficial mutations are available, natural selection must purge deleterious variants and amplify beneficial mutations simultaneously.  These forces can conflict with each other in closely linked regions of the genome, leading to a second source of stochasticity known as \emph{genetic draft} \citep{gillespie:2000}. Thus, provided that the cost of a deleterious mutation is not too high, it can hitchhike to high frequency with a beneficial ``driver'' mutation that happens to arise on the same genetic background \citep{maynard-smith:haigh:1974}. The fixation of these deleterious ``passengers'' imposes a direct cost on the fitness of the population, which can only be ameliorated by future compensatory mutations. Deleterious mutants also impose an opportunity cost on the fitness of the population when they hinder the fixation of driver mutations that arise on poor genetic backgrounds \citep{charlesworth:1994, peck:1994}. Thus, even when they are not destined to fix, segregating deleterious variants still contribute to an overall mutation load, which reduces the fraction of available genetic backgrounds where adaptation can proceed unhindered. 

Together, deleterious passengers and the mutation load can dramatically reduce the rate of adaptation, and in extreme cases, even lead to fitness decline \citep{silander:etal:2007} and mutational meltdown \cite{gabriel:etal:1993}. Conversely, even a small number of beneficial mutations will bias the spectrum of deleterious mutations that accumulate during the course of evolution \citep{schiffels:etal:2011}. These interactions between adaptation and constraint have been the subject of extensive theoretical study \citep{charlesworth:1994, peck:1994, barton:1995, orr:2000, johnson:barton:2002, bachtrog:gordo:2004, desai:etal:2007, jiang:etal:2011, hartfield:otto:2011, schiffels:etal:2011, goyal:etal:2012, mcfarland:etal:2014}, but many aspects of this process remain poorly characterized. In particular, theory still struggles to account for observed variation in the fitness effects of new mutations, and how these disparate mutations combine to determine overall levels of hitchhiking and the genetic load. This gap in our understanding is especially problematic for the largest and most rapidly adapting populations, where multiple beneficial driver mutations compete for fixation at the same time. As we will see, these populations actually accumulate \emph{more} deleterious mutations, even as they become more efficient at finding and fixing adaptive variants. It is therefore unsurprising that deleterious passengers are thought to play an important role in the adaptive process \citep{pybus:etal:2007, mcfarland:etal:2013, covert:etal:2013, luksza:laessig:2014}. 

In this article, we study the effects of deleterious passengers in a simple model of widespread adaptation. We employ a perturbative approach, leveraging a recent mathematical description of adaptation in the absence of deleterious mutations \citep{good:etal:2012, fisher:2013}. This enables us to obtain simple analytical predictions for genomic substitution rates across a broad range of beneficial and deleterious fitness effects. In particular, we find that the maximum cost of a passenger is inversely proportional to the coalescence timescale, $\tc$, over which the fates of new common ancestors are determined. This constitutes a natural generalization of the traditional drift-barrier in the presence of widespread genetic draft, where the population-size dependence of $\tc$ can be dramatically altered. We end by discussing the relevance of these findings for recent microbial evolution experiments and comment on directions for future work.

\section*{Model}

\noindent We consider a population of $N$ nonrecombining haploid individuals that accumulate mutations at a per genome rate $U$. We assume an infinite sites model, in which the fitness effect of each mutation is drawn from a distribution of fitness effects, $\rho(s)$, that remains constant over the relevant time interval. We further partition the distribution of fitness effects (the DFE) into its beneficial and deleterious components,
\begin{align}
\label{eq:dfe-partition}
U \rho(s) = \begin{cases}
\Ub \rhob(s) & \text{if $s > 0$,} \\
\Ud \rhod(-s) & \text{if $s < 0$,}
\end{cases}
\end{align}
where $\Ub$ and $\Ud$ denote the per-genome rates of beneficial and deleterious mutations, respectively. 
For concreteness, we will primarily focus on a simplified ``two-effect'' DFE,
\begin{align}
\label{eq:two-effect-model}
U \rho(s) = \Ub \delta(s-\sb) + U_d \delta(s+\sd) \, ,
\end{align}
where beneficial and deleterious mutations each have a characteristic fitness effect. In a later section, we will show how our analysis can be extended to more general distributions, provided that $\Ub \rhob(s)$ and $\Ud \rhod(s)$ satisfy certain technical conditions. 

These assumptions define a simple model of sequence evolution with a straightforward computational implementation. We wish to use this model to study the impact of deleterious mutations on the long-term genetic composition of the population, which is determined by the average substitution rate, $R(s)$, of new mutations as a function of their fitness effect. In particular, we wish to quantify the relative contributions from beneficial and deleterious mutations. For the simple two-effect DFE in \eq{eq:two-effect-model}, this is uniquely determined by the total beneficial and deleterious substitution rates, $\Rb = R(\sb)$ and $\Rd = R(-\sd)$. However, for more general DFEs, there is some ambiguity in how we define the net contribution from beneficial and deleterious mutations. For example, the raw substitution rates $\int_0^\infty R(s) \, ds$ and $\int_0^\infty R(-s) \, ds$ tend to be dominated by neutral or nearly-neutral mutations, which have a negligible impact on the fitness of the population. To avoid this bias, we focus on the \emph{weighted} substitution rates,
\begin{align}
\label{eq:substitution-rate-definition}
\Rb =  \int_0^\infty  \frac{s}{\sbbar} \cdot R(s) \, ds \, , \quad \Rd = \int_0^\infty \frac{s}{\sdbar} \cdot R(-s) \, ds \, ,
\end{align}
where $\sbbar = \int_0^\infty s \rhob(s) \, ds$ and $\sdbar = \int_0^\infty s \rhod(s) \, ds$ represent the average fitness effects of the underlying DFE. For the two-effect DFE in \eq{eq:two-effect-model}, $\Rb$ and $\Rd$ coincide with the raw rates of sequence evolution, as desired. For more general DFEs, the substitution rate of each mutation is weighted by its contribution to the total fitness of the population, so that the total rate of adaptation is simply $v = \sbbar \Rb - \sdbar \Rd$.

\section*{Heuristic Analysis and Intuition}

\noindent Before we perform any explicit calculations, it will be useful to consider the dynamics of deleterious mutations from a heuristic perspective. This will allow us to identify many of the relevant fitness scales, and will help build intuition for the more detailed calculations below. Our discussion will resemble the traditional ``drift-barrier'' argument from the introduction, but it will apply for a much broader range of populations where drift is no longer the dominant evolutionary force. 

\begin{figure}
\includegraphics[width=0.95\columnwidth]{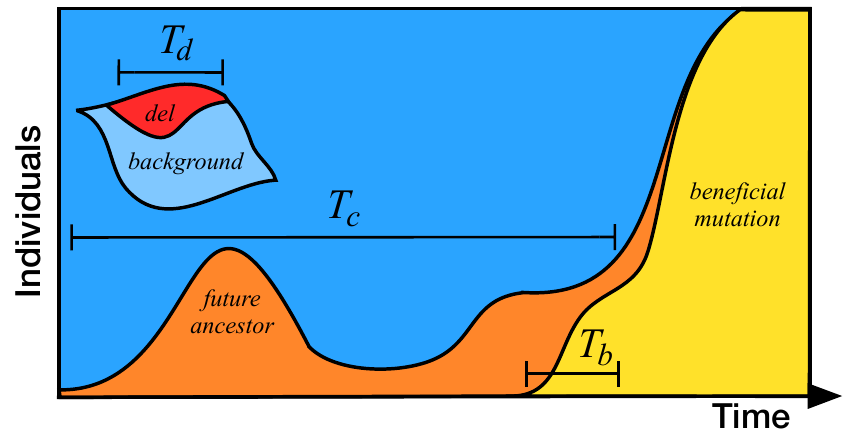}
\caption{A schematic depiction of the fundamental evolutionary timescales. A deleterious variant ($del$) is purged relative to its background on a characteristic timescale $\td$. Meanwhile, the fate of a future common ancestor is determined on a characteristic timescale $\tc$, and the fate of a beneficial mutation is determined over a characteristic timescale $\tb$. \label{fig:heuristic-diagram}}
\end{figure}

Deleterious mutations can be classified into two fundamental regimes depending on the substitution rates of the mutations involved. Sufficiently weakly selected mutations will accumulate nearly neutrally ($\Rd \approx \Ud$), while sufficiently strongly selected mutations will rarely fix ($\Rd \approx 0$). The transition between these two regimes will occur for some characteristic cost $\sdc$, which can be estimated from the fundamental timescales of the system. When a deleterious mutation arises, it originates on a particular genetic background, and it competes with this background lineage until one of them is driven to extinction. If the fitness of the background is sufficiently high, it is possible for the deleterious mutation to increase in frequency in the short term. Yet on average, the frequency of the mutant relative to its background will decay exponentially at rate $\sd$ (see \fig{fig:heuristic-diagram}). Thus, deleterious mutations are typically purged by selection on a characteristic timescale $\td \sim \sd^{-1}$. 

The fixation of these mutations is governed by the underlying coalescent process. At each site in the genome, exactly one of the present-day individuals will grow to become an ancestor to the entire population. We let $\tc$ denote the characteristic timescale over which the fates of new common ancestors are determined. This does not imply that new common ancestors have fixed within $\tc$ generations, but only that their chances of extinction are negligible beyond this point (see \fig{fig:heuristic-diagram}). As we will demonstrate below, $\tc$ is closely related to the \emph{coalescent timescale} that determines the levels of neutral diversity in the population. When the genealogy of the population is dominated by drift, $\tc$ is simply proportional to the population size ($\tc \approx 2N$), but in general $\tc$ can vary in an arbitrary way with the underlying parameters. A deleterious mutation can only fix if it arises in a future common ancestor and evades natural selection for $\tc$ generations. If $\td \ll \tc$, selection will typically purge the deleterious variant before its descendants reach fixation, while mutations with $\td \gg \tc$ will barely feel the effects of selection before they fix. This implies that the crossover between effectively neutral ($\Rd \approx \Ud$) and effectively lethal ($\Rd \approx 0$) substitution rates must occur for $\sdc = c \tc^{-1}$, where $c$ is an $\mathcal{O}(1)$ constant. 

So far, we have considered two classes of deleterious mutations: those that fix and those that do not, with a transition between the two regimes at $\sdc \sim \tc^{-1}$. In an adapting population, there is also a third class of deleterious mutations: those that are most likely to hinder the spread of the beneficial variants, regardless of whether or not they fix. The typical cost of these maximally interfering mutations can be estimated from a similar timescale argument. Here, the relevant timescale is not $\tc$ but rather the characteristic time $\tb$ over which the fate of a \emph{beneficial} mutation is determined (see \fig{fig:heuristic-diagram}). When the fates of beneficial mutations are controlled by genetic drift, $\tb$ is simply the drift time of the beneficial mutation ($\tb \sim 1/\sb$), but in general $\tb$ can be an arbitrary function of the underlying parameters. For a deleterious mutation to hinder the fixation of a beneficial variant, it must fix (or nearly fix) within the beneficial lineage \emph{and} begin to feel the effects of selection within $\tb$ generations. These conditions are jointly satisfied when $\tb \sim \td$, so the maximally interfering mutations have $\sdb = b \tb^{-1}$, where $b$ is another $\mathcal{O}(1)$ constant. 

Thus, without performing any explicit calculations, we see that a simple heuristic argument is sufficient to determine the relevant deleterious fitness effects in terms of the fundamental timescales of the system. In the following sections, we will rederive these results more rigorously with explicit calculations of $\Rb$ and $\Rd$ in several different parameter regimes. Although these calculations are somewhat less general than the heuristic argument above, they will allow us to predict the \emph{quantitative} nature of the transitions near $\sdc$ and $\sdb$ in addition to the location of the transitions themselves. Perhaps more importantly, these calculations provide explicit expressions for $\tc$ and $\tb$ in terms of the underlying parameters $N$ and $U\rho(s)$, which enables us to estimate when deleterious passengers are likely to be important in practice. 

\section*{Analysis} 

\noindent Although our model is simple, it can be difficult to model the evolution of a tightly linked genome directly at the sequence level. The fate of any particular variant is strongly influenced by additional mutations (particularly beneficial driver mutations) that segregate in the same genetic background, as well as competing mutations that arise elsewhere in the population. Keeping track of the arrival times and haplotype structure of these mutations can rapidly become unwieldy when the genome contains more than a handful of selected sites. 

Fortunately, previous work has shown that many of these difficulties can be avoided by utilizing an intermediate level of description, where the distribution of fitnesses within the population plays a central role \citep{haigh:1978, tsimring:etal:1996}. Instead of tracking individual genotypes, we focus on the total fraction of the population, $f(X,t)$, with (log) fitness $X$. We can then write down a consistent set of equations governing the evolution of the fitness distribution without reference to the underlying genotypes. Similarly, the fate of a new mutation can be recast as a competition between the fitness distribution of its descendants and that of the background population (see \app{appendix:stochastic-model}). This leads to a dramatic simplification in large populations, since the distribution of fitnesses can be highly predictable even when sequence evolution is highly stochastic. 

The dynamics of the fitness distribution and the fates of individual mutations have been well-characterized in the absence of deleterious mutations \citep{neher:etal:2010, hallatschek:2011, good:etal:2012, fisher:2013}. Thus, rather than looking for an exact solution in the deleterious case, we will utilize a \emph{perturbative} approach, focusing only on the leading order corrections to the substitution rate in the limit that $\Ud \to 0$. This strategy allows us to exploit our existing knowledge of the evolutionary dynamics in the absence of deleterious mutations. In particular, it implies that the fundamental fitness scales $\sdc \sim \tc^{-1}$ and $\sdb \sim \tb^{-1}$ can be estimated from previously derived formulae in the $\Ud \to 0$ limit. Although this limit may seem unrealistic (given that deleterious mutations are at least as common as their beneficial counterparts), these leading-order expressions will turn out to be surprisingly accurate in large populations, even when $\Ud \gg \Ub$. In the following analysis, we will distinguish between different regimes depending on the frequency of strongly beneficial ``driver'' mutations, which can dramatically influence the timescales $\tc$ and $\tb$.

\subsection*{No driver mutations}

\noindent We start by reviewing the simplest case, where beneficial driver mutations can be neglected. This assumption applies not only when $\Ub=0$, but also in small populations where drivers fix less frequently than neutral coalescence events ($N\Rb \ll 1$). Both conditions are sufficient to ensure that the coalescent timescale is dominated by genetic drift ($\tc \sim N$). In the absence of deleterious mutations, there is no fitness diversity within the population [i.e., $f_0(X,t) \approx \delta(X)$]. To leading order, the fate of a particular individual is determined solely by the balance between selection and drift, and the fixation probability is given by the standard formula 
\begin{align}
\pfix(X) \approx \frac{2X}{1-e^{-2NX}} + \mathcal{O}(\Ud) \, ,
\end{align}
where $X$ denotes the fitness of the individual \citep{fisher:1930, wright:1931}. The deleterious substitution rate trivially follows by averaging over the potential fitness backgrounds of new deleterious mutations:
\begin{align}
\label{eq:single-locus-rd}
\Rd = N\Ud \int f_0(X) \pfix(X-\sd) \, dX \approx \frac{2 N \Ud \sd}{e^{2 N \sd} - 1 } \, .
\end{align}
Thus, we recover the well-known result that deleterious mutations accumulate neutrally when $\sd \ll 1/2N$, and are exponentially suppressed when $\sd \gg 1/2N$. As we argued on heuristic grounds above, the transition between these two extremes occurs at $\sdc \approx 1/2N$, when the lifetime of a deleterious mutation ($\tsel \sim 1/\sd$) is on the order of the neutral coalescence time ($\tc \sim N$).
 
\begin{figure}
\centering
\includegraphics[width=0.9\columnwidth]{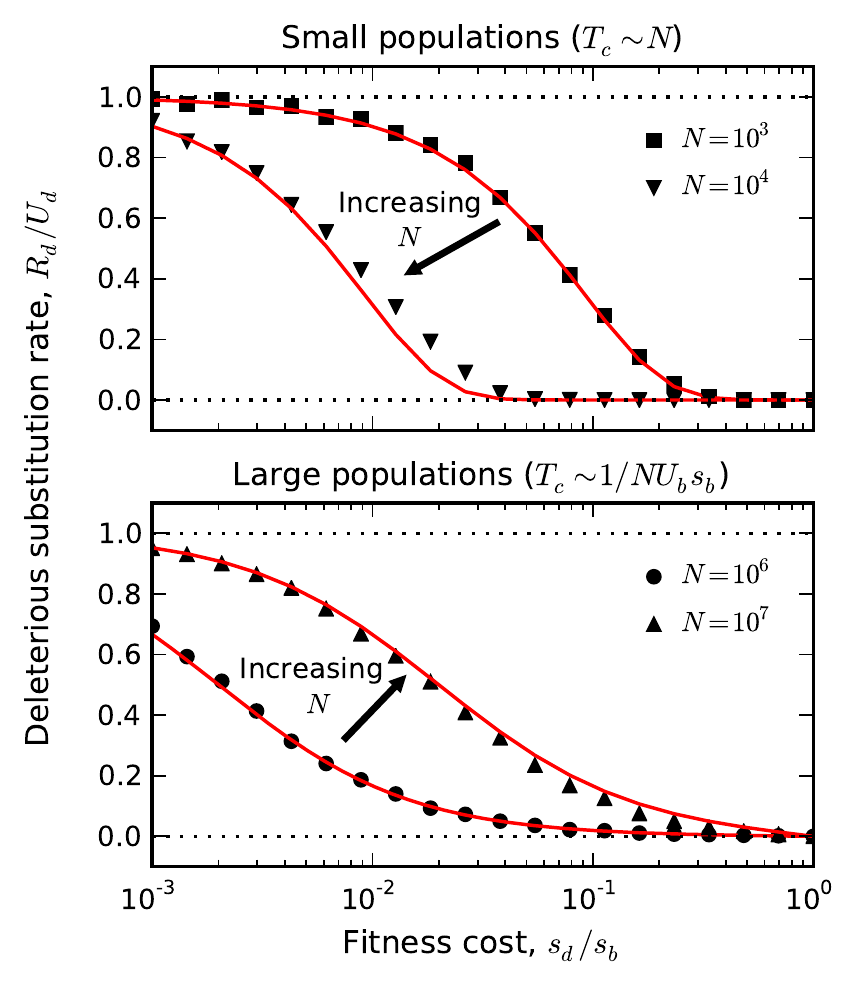}
\caption{The deleterious substitution rate in the limit that beneficial driver mutations are rare. Symbols represent simulations of a two-effect DFE for fixed $\Ub=10^{-9}$, $\sb=10^{-2}$, and $\Ud=10^{-4}$, with $\sd$ tuned between $10^{-5}$ and $10^{-2}$. In the top panel, the red lines give the drift-dominated predictions from \eq{eq:single-locus-rd}, while the bottom panel shows the sweep -dominated predictions from \eq{eq:sweep-rd}. \label{fig:sweep-rd}}
\end{figure}

Of course, the substitution rate in \eq{eq:single-locus-rd} eventually breaks down for large values of $\Ud$, when many deleterious mutations segregate in the population simultaneously. Interference between these mutations can drive additional mutants to fixation via Muller's ratchet \citep{muller:1964}, thereby accelerating the rate of sequence evolution. Yet while the ratchet can alter both the functional form of \eq{eq:single-locus-rd} and the location of the transition to neutrality, it preserves the monotone dependence of these quantities on the population size (see \app{appendix:first-order}). Thus, in the absence of drivers, both the number of deleterious substitutions and the fitness effects of these mutations tend to decrease in larger populations (\fig{fig:sweep-rd}A). 

In the following sections, we will assume for simplicity that the population size is large enough that none of the deleterious mutations would fix due to drift alone ($N\sd \gg 1$). This is a reasonable assumption for most microbial evolution experiments, where the effective population sizes are on the order of $10^5$ or greater \citep{kawecki:etal:2012}. This also allows us to focus on true ``passenger'' mutations, which could never fix without the help of a beneficial driver. [For an analysis of the opposite regime, see \citet{mcfarland:etal:2014}.] As we demonstrate below, these passenger mutations display a qualitatively different dependence on the population size than \eq{eq:single-locus-rd} would predict. 

\subsection*{Rare driver mutations}

\noindent In larger populations, beneficial drivers substitute sufficiently often that the coalescence timescale is dominated by the sweep time of a beneficial mutation. In this case, we still expect deleterious mutations to accumulate neutrally when $\sd \ll \tc^{-1}$, but this threshold is no longer tied to the inverse population size. Again, it is useful to begin with the simplest case, where beneficial mutations are sufficiently rare that they fix independently. This requires that the waiting time for a successful driver [$\twait \approx 1/(2N\Ub\sb)$] is much longer than its fixation time [$\tfix \approx \frac{2}{\sb} \log(2N\sb)$], so that $\tc \sim \twait$. The effects of deleterious mutations in this ``rare driver'' regime have been studied by a number of previous authors \citep{charlesworth:1994, peck:1994, orr:2000, johnson:barton:2002}. This earlier work primarily focuses on the reduction in the beneficial substitution rate, with analytical results available in the limiting case where $\sd > \sb$. Here, we generalize these results and derive simple analytical formulae for $\Rb$ and $\Rd$ which are valid across the full range of deleterious fitness costs. These formulae can then be contrasted with their counterparts in the multiple driver regime below. 

\begin{figure}
\includegraphics[width=0.9\columnwidth]{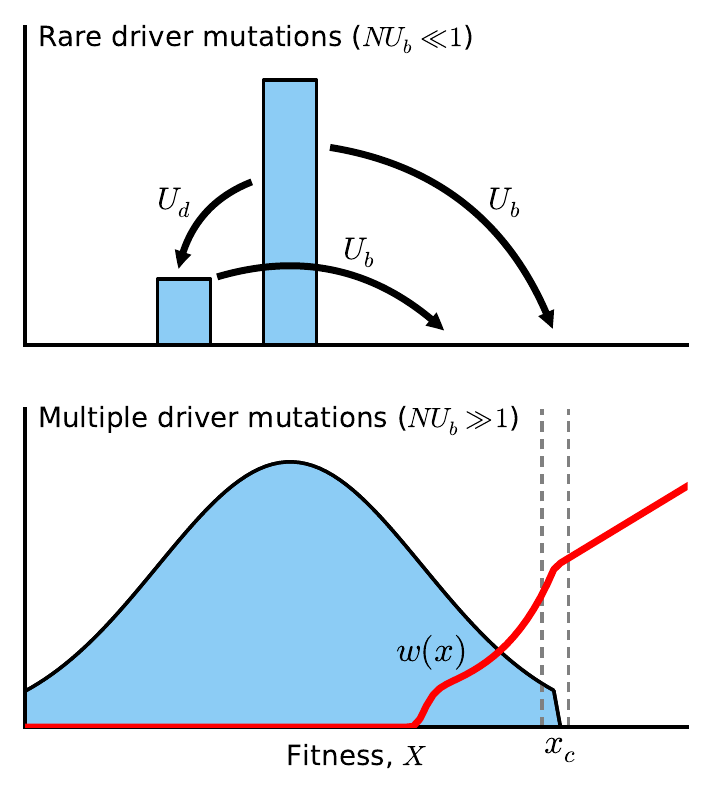}
\caption{A schematic illustration of the fitness distribution. In the rare driver regime (top), the population is predominantly composed of the last successful driver, with a deleterious subpopulation given by \eq{eq:sweep-f}. In the multiple driver regime (bottom), the fitness distribution approaches a steady-state shape, $f(x)$, that translates towards higher fitness at rate $v$. The red line depicts the fixation probability, $w(x)$, which increases rapidly with $x$ before transitioning to the standard Haldane result at $x\approx \xc$. \label{fig:diagram}}
\end{figure}

In the absence of deleterious mutations, a population in the rare driver regime is typically fixed for a single genotype (i.e., the last successful driver) whose fitness can be taken to be $X=0$. Deleterious mutations create variation around this genotype, which is temporarily depleted when the next driver sweeps to fixation (\fig{fig:diagram}). If $t$ denotes the time since the fixation of the last driver, then the distribution of fitnesses in the population is given by
\begin{align}
\label{eq:sweep-f}
f(X,t) \approx \begin{cases}
1-\frac{\Ud}{\sd} \left(1-e^{-\sd t} \right) & \text{if $X = 0$,} \\
\frac{\Ud}{\sd} \left( 1 - e^{-\sd t} \right) & \text{if $X = -\sd$,} \\
0 & \text{else,}
\end{cases} 
\end{align}
where we have retained only the leading order contribution in $\Ud$ \citep{johnson:1999}. At long times ($t \gg 1/\sd$), this distribution approaches the standard mutation-selection balance, $f(-\sd) \approx \Ud/\sd$, but it is possible that the next driver will occur before this equilibrium is reached \citep{johnson:barton:2002}. The waiting time for the next successful driver is exponentially distributed with mean $\tc \approx 1/(2N\Ub\sb)$, so the average fraction of deleterious individuals at the time of the next sweep is
\begin{align}
\label{eq:average-sweep-f}
f(-\sd) = \frac{\Ud}{2N\Ub\sb+\sd} = \frac{\Ud}{\sd} \left(\frac{\tc \sd}{1+\tc \sd}\right) \, ,
\end{align}
which reduces to mutation-selection balance when $\sd \gg \tc^{-1}$. Previous studies often neglect this relaxation phase, since they focus on deleterious fitness effects with $\sd \gtrsim \sb \gg \tc^{-1}$. But based on our heuristic discussion, we expect that most of the successful passenger mutations will have $\sd \lesssim \tc^{-1}$, where the deviations from mutation-selection balance in \eq{eq:average-sweep-f} start to become important. Indeed, as we will see below, it is exactly this time-dependent behavior that drives most of the deleterious hitchhiking in these populations.   

When the next driver mutation does arise, it can drag a deleterious passenger to fixation in one of two ways. The deleterious mutation can arise before the driver mutation, so that the driver originates directly in a deleterious background. These passenger-first events occur at rate $N\Ub f(-\sd) \pfix(\sb-\sd)$. Alternatively, the deleterious mutation can arise after the driver and fix within the driver lineage while it is still rare. In \app{appendix:first-order}, we show that these driver-first events occur at rate $N\Ub \frac{\Ud}{\sb} \pfix(\sb-\sd)$. When $\sd \sim \tc^{-1}$, this rate is much smaller than the passenger-first scenario above, but it becomes comparable in magnitude when $\sd \sim \sb$. Combining these two expressions, we find that the total deleterious substitution rate is given by
\begin{align}
\label{eq:sweep-rd}
\Rd & \approx \begin{cases} 
\frac{\Ud \left[ 1 - \left( \frac{\sd}{\sb} \right)^2 \right] }{ 1 + \left( \frac{\sd}{2N\Ub\sb} \right) } & \text{if $\sd < \sb$,} \\
0 & \text{else.} 
\end{cases}
\end{align}
A similar expression was derived by \citet{schiffels:etal:2011} using a different method of analysis. The substitution rate in \eq{eq:sweep-rd} approaches the neutral limit when $\sd \ll 2N\Ub\sb$ and decays as a power law when $\sd \gg 2N\Ub\sb$. Thus, even when deleterious mutations would never drift to fixation on their own, frequent drivers can still cause these variants to accumulate like neutral mutations. Note that the time-dependent fitness distribution in \eq{eq:sweep-f} played a crucial role in the emergence of this effectively neutral regime. Once the fitness distribution has reached mutation-selection balance, hitchhiking is already reduced by a factor of $N\Ub \ll 1$. In agreement with our heuristic argument, the border of the effectively neutral regime is located at $\sdc \sim 2N\Ub\sb$, when the lifetime of a deleterious mutation ($\tsel \sim 1/\sd$) is on the order of the coalescence time ($\tc \sim \twait$). However, in contrast to the non-adapting case, $\sd^*$ is now an \emph{increasing} function of the population size. This implies that \emph{more} deleterious mutations will accumulate in larger and more rapidly adapting populations, and that the average cost of each passenger will increase as well (\fig{fig:sweep-rd}B). 

To calculate the total rate of sequence evolution and the rate of adaptation, we must also understand how deleterious mutations influence the fixation of the drivers. In the absence of deleterious mutations, drivers substitute at rate $\Rb=2N\Ub\sb$, and deleterious mutations will reduce this rate by decreasing the effective fitness advantage of the drivers. Part of the reduction in driver fitness arises from the fixation of deleterious passengers within the driver lineage before it completes its sweep. This can occur via either of the two hitchhiking scenarios (passenger-first hitchhiking or driver-first hitchhiking) described above. Deleterious mutations can also influence the fates of \emph{unloaded} drivers through the time-varying mean fitness in \eq{eq:sweep-f}. The corresponding reduction in $\Rb$ is somewhat more difficult to obtain compared to the deleterious substitution rate, since $\Rb$ depends on non-equilibrium properties of $f(X,t)$ that are not captured by the simple average in \eq{eq:average-sweep-f}. Nevertheless, since the fate of a driver mutation is determined while it is at low frequency, the reduction in $\Rb$ can still be obtained using standard branching-process techniques \citep{johnson:barton:2002}. We carry out this calculation in \app{appendix:first-order} and find that the leading order reduction in $\Rb$ is given by
\begin{align}
\label{eq:sweep-rb}
\Rb = \begin{cases}
2N\Ub \left( \sb - \frac{\Ud \sd}{\sb} \cdot \frac{\sd}{2N\Ub\sb+\sd} \right) & \text{if $\sd < \sb$,} \\
2N \left( 1 - \frac{\Ud}{\sd} \right) \Ub \sb & \text{else.}
\end{cases}
\end{align}
Most of the interesting reduction occurs for $\sd \gg \tc^{-1}$, when the fractional change in $\Rb$ becomes independent of $\tc$. In other words, these deleterious mutations have all reached mutation-selection balance by the time that the next driver arises. When $\sd < \sb$, \eq{eq:sweep-rb} has a simple interpretation as a reduction in the establishment probability of each driver due to the deleterious passengers that accumulate during the establishment time, $\test \sim \sb^{-1}$. These ``tunneling'' events are crucial for obtaining the proper $\sd$ dependence in \eq{eq:sweep-rb}; traditional arguments based on the mutation load \citep{kimura:maruyama:1966} would otherwise suggest that $\Rb$ is \emph{independent} of $\sd$. In the opposite case where $\sd > \sb$, we recover the well-known ``background selection'' \citep{charlesworth:1994} or ``ruby in the rough'' \citep{peck:1994} behavior observed in previous studies. In this case, loaded drivers can never fix, and \eq{eq:sweep-rb} can be interpreted as a reduction in the effective population size equal to the fraction of mutation-free individuals in the population.  

These results show that the largest reduction in $\Rb$ occurs for $\sd = \sb$, which is much larger than the size of a typical passenger mutation ($\sdc \sim 2N\Ub\sb$). In other words, the maximally interfering mutations rarely hitchhike to fixation ($\Rd \approx 0$). The disparity between these two scales can be explained in terms of the heuristic argument above. When drivers are rare, the fates of beneficial mutations are primarily influenced by drift. The driver fates are therefore determined during the drift time, $\tb \approx \sb^{-1}$, which is much shorter than both the fixation time of a driver [$\tfix=\frac{2}{\sb} \log( 2 N \sb)$] and the waiting time between sweeps [$\twait \sim 1/(2N\Ub\sb)$]. This will change dramatically in the multiple driver regime below.

\subsection*{Multiple driver mutations}

\noindent As the population size increases, the waiting time between drivers is eventually dwarfed by the time that it takes each driver to fix. Multiple drivers will segregate in the population at the same time, and these mutations will interfere with each other as they compete for fixation. In this \emph{clonal interference} regime, drivers and passengers are \emph{both} dominated by the effects of genetic draft, and coalitions of multiple drivers are often required to drive a lineage to fixation \citep{rouzine:etal:2003, desai:fisher:2007}. Recent empirical work in microbial populations suggests that this regime is likely to be the rule rather than the exception \citep{miralles:etal:1999, deVisser:etal:1999, parfeito:etal:2007, miller:etal:2011, batorsky:etal:2011, strelkowa:laessig:2012, lang:etal:2013, lee:marx:2013, kvitek:sherlock:2013, barroso-batista:etal:2014}, so it is important that we extend our previous analysis to this potentially more realistic scenario. 

At long times, a population in the multiple driver regime reaches a steady-state in which the continuous production of new mutations is balanced by the depletion of this diversity due to natural selection \citep{tsimring:etal:1996, rouzine:etal:2003, desai:fisher:2007}. The fitness distribution $f(X,t)$ behaves like a ``traveling wave,'' with a characteristic shape $f(x)$ that translates towards higher fitness at a constant rate $v$ (see \fig{fig:diagram}). In some respects, this multi-driver equilibrium is simpler to analyze than the rare-driver regime above, where the punctuated nature of adaptation required us to explicitly account for departures from steady-state. Averaged over short fitness and time scales, the steady-state shape of the fitness distribution is described by the deterministic dynamics, 
\begin{align}
\label{eq:f-equation}
- v \frac{\partial f}{\partial x} = x f + U \int ds \, \rho(s) \left[ f(x-s) - f(x) \right] \, ,
\end{align}
where $x$ denotes the \emph{relative fitness}, $X - \overline{X}(t)$. This deterministic equation holds throughout the bulk of the fitness distribution, but starts to break down near the high-fitness ``nose'' (see \fig{fig:diagram}) where most successful drivers originate. In large populations, the nose constitutes just a small fraction of the total population, so we can approximate the effects of genetic drift in this regime with a suitable linear branching process \citep{neher:etal:2010, good:etal:2012, fisher:2013}. The fixation probability, $w(x)$, for a lineage with relative fitness $x$ satisfies a related differential equation,
\begin{align}
\label{eq:w-equation}
v \frac{\partial w}{\partial x} = x w + U \int ds \, \rho(s) \left[ w(x+s) - w(x) \right] - \frac{w^2}{2} \, ,
\end{align}
where the nonlinear term gives the contribution from genetic drift (see \app{appendix:stochastic-model}). The marginal fixation probability, $\pfix(s)$, of a mutation with fitness effect $s$ can be calculated from $w(x)$ by averaging over the distribution of background fitnesses, 
\begin{align}
\label{eq:pi-equation}
\pfix(s) = \int f(x) w(x+s) \, dx \, ,
\end{align}
where the consistency condition $\pfix(0) \approx 1/N$ serves to uniquely determine $v$ as a function of the underlying parameters $N$ and $U \rho(s)$ \citep{hallatschek:2011, good:etal:2012, fisher:2013}.  

In the absence of deleterious mutations, we have previously derived an approximate solution to \consecutiveeqs{eq:f-equation}{eq:pi-equation} in the strong selection regime where $\sb \gg \sqrt{v}$ \citep{good:etal:2012}. Unfortunately, this solution contains several pathologies that render it unsuitable for the perturbative analysis below \citep{fisher:2013}. In \app{appendix:zeroth-order}, we derive a modified version of this solution that corrects these issues. The resulting fixation probability is characterized by a narrow boundary layer near a critical fitness value, $\xc \geq \sb$. Above this point, lineages fix without the need for additional driver mutations, so $\wo(x) \approx 2x$. Below $\xc$, the fixation probability rapidly declines as
\begin{align}
\label{eq:approx-w}
\wo(x) \approx \begin{cases}
2 \xc e^{\frac{x^2-\xco^2}{2\vo}} & \text{if $x>\xc-\sb$,} \\
0 & \text{else.}
\end{cases}
\end{align}
The fitness distribution displays a similar transition near $\xc$, taking a simple Gaussian form below $\xc$,
\begin{align}
\label{eq:approx-f}
\fo(x) \approx \frac{1}{\sqrt{2 \pi \vo}} e^{-\frac{x^2}{2\vo}} \, ,
\end{align} 
and vanishing above this threshold \citep{fisher:2013}. The location of the boundary can be obtained from an integral transform of \eq{eq:w-equation}, which yields an auxiliary condition
\begin{align}
\label{eq:auxilliary-condition}
1 = \frac{\Ub}{\sb} \left[ 1 - \frac{\sb}{\xc} \right]^{-1} e^{\frac{\xc \sb}{\vo} - \frac{{\sb}^2}{2\vo}} \, ,
\end{align}
which uniquely determines $\xc$ as a function of $\Ub$, $\sb$, and $\vo$ \citep{good:etal:2012}. 

The solution in \consecutiveeqs{eq:approx-w}{eq:auxilliary-condition} requires that $(\xc-\sb) \gg \sqrt{v}$ and $\sb \gg \sqrt{v}$. The first of these conditions places a lower bound on the amount of clonal interference in the population: there must be sufficient fitness variation that the \emph{parents} of successful drivers have abnormally high fitness (i.e., greater than one standard deviation from the mean). This distinguishes the clonal interference regime from the rare driver limit above. The second condition places an upper bound on the amount of fitness diversity in the population: individuals that comprise the bulk of the population (i.e., within one standard deviation from the mean) typically harbor the same number of driver mutations. In terms of the underlying parameters, these two conditions require that $N\sb \gg 1$ and $\Ub \ll \sb$ \citep{desai:fisher:2007}. We focus on this regime because it is thought to apply to a broad range of microbial evolution experiments, at least in the initial phases of adaptation \citep{desai:etal:2007, parfeito:etal:2007, wiser:etal:2013, barroso-batista:etal:2014}. 

\begin{figure}
\centering
\includegraphics[width=0.9\columnwidth]{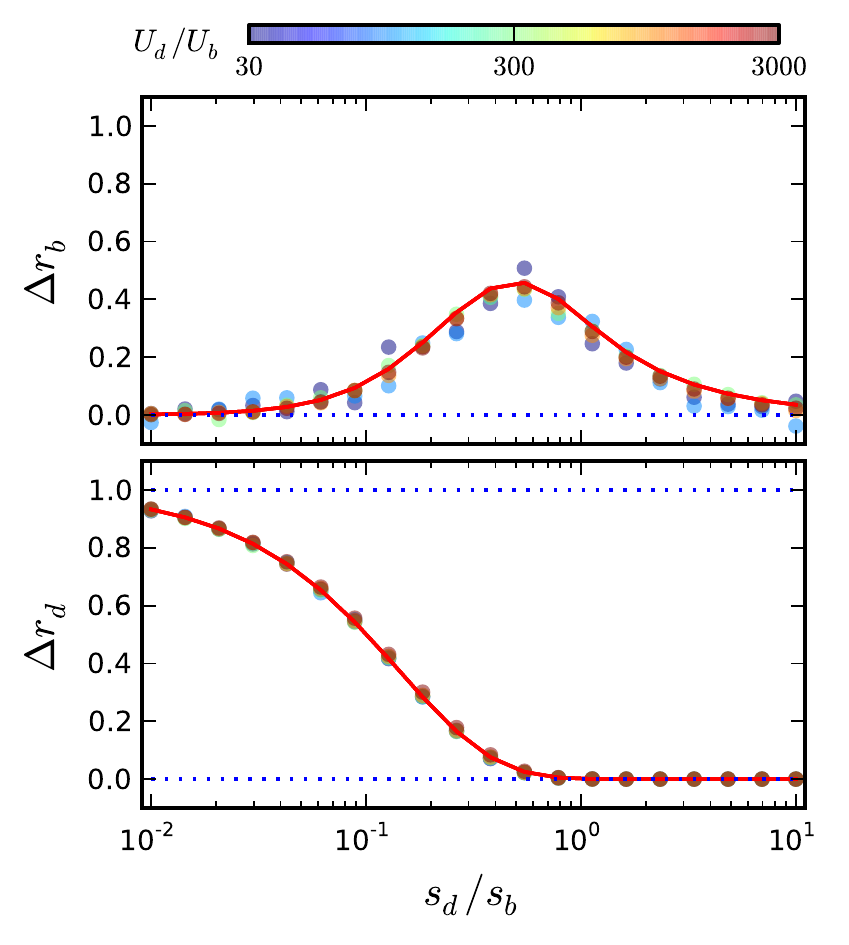}
\caption{Corrections to the substitution rate in the limit that driver mutations are common. Symbols denote the results of forward-time simulations for various combinations of $\Ud$ and $\sd$, with the remaining parameters fixed at $N=10^7$, $\Ub=10^{-5}$, $\sb=10^{-2}$. The quantities $\etarb$ and $\etard$ are calculated from \eq{eq:perturbative-expansions}, with $\vo$ measured from simulations and $\xc$ estimated from \eq{eq:auxilliary-condition}. For comparison, the solid red lines show the first-order predictions from \eqs{eq:full-etard}{eq:full-etarb}, using the same estimated values of $\vo$ and $\xc$. \label{fig:delta-collapse}}
\end{figure}

Deleterious mutations lead to a reduction in the fixation probability and deviations from the Gaussian fitness distribution, which can alter the rate of adaptation and the location of the nose in potentially complex ways. However, we can still investigate the leading order effects of deleterious passengers using the same perturbative strategy that we employed above. We rewrite the substitution rates in the suggestive form: 
\begin{subequations}
\label{eq:perturbative-expansions} 
\begin{align}
\Rb & = \Rbo \left[ 1 - \left( \frac{\Ud}{\xc} \right) \etarb \right] \, , \\
\Rd & = 0 + \Ud \cdot  \etard \, , 
\end{align}
with corresponding expansions for $w(x)$, $f(x)$, and $v$:
\begin{align} 
& w(x) = \wo(x) \left[ 1 + \left( \frac{\Ud}{\xc} \right) g(x) \right] \, , \\
& f(x) \propto \fo(x) \left[ 1 + \left( \frac{\Ud}{\xc} \right) h(x) \right] \, , \\
& v = \vo \left[1 - \left( \frac{\Ud}{\xc} \right) \left( \etarb + \frac{\xc \sd}{\vo} \etard \right) \right] \, . 
\end{align}
\end{subequations}
Note that we have not included an expansion for $\xc$, since this is simply a property of $\wo(x)$ rather than a measurable quantity like $v$. Thus, with a slight abuse of notation, we will continue to use $\xc$ to denote the zeroth-order value $\xco$. With these definitions in hand, we can substitute \eq{eq:perturbative-expansions} into \consecutiveeqs{eq:f-equation}{eq:pi-equation} and equate like powers of $\Ud$ to obtain at a corresponding set of equations for $\etarb$, $\etard$, $g(x)$ and $h(x)$ (see \app{appendix:first-order}). The correction to the deleterious substitution rate is particularly easy to calculate, since the zeroth order ($U_d = 0$) contribution vanishes. To leading order, we find that 
\begin{align}
\etard & = N \int \fo(x) \wo(x-\sd) \, dx \, , \nonumber \\
    & \approx \left( \frac{ 1 - e^{-\frac{\sb \sd}{\vo}}}{\frac{\sb \sd}{\vo}} \right) \exp \left[ -\frac{(\xc-\sb) \sd}{\vo} \right] \, , \label{eq:full-etard} 
\end{align}
which interpolates between the effectively neutral limit ($\Rd\approx\Ud$) and the effectively lethal limit ($\Rd\approx0$) illustrated in \fig{fig:delta-collapse}. In sufficiently large populations where $\xc \gg \sb$, \eq{eq:full-etard} reduces to
\begin{align}
\label{eq:approx-etard}
\etard \approx \exp \left[ - \frac{\xc \sd}{\vo} \right] \, ,
\end{align}
which shows that the border of the effectively neutral regime occurs at $\sdc \sim \vo/\xc$. This crossover has a natural interpretation in terms of the fundamental timescales of the system. The ``nose-to-mean'' time $\tsw \sim \xc/\vo$ is the time required for the current fitness of the nose to become the mean fitness of the population. When $\xc \gg \sb$, we have previously shown that this is also the timescale over which the fates of new common ancestors and successful drivers are decided, so that $\tc \sim \tb \sim \tsw$ \citep{desai:etal:2013}. Solving for $\vo$ and $\xc$, one can show that  
\begin{align}
\tsw \approx \frac{1}{\sb} \log \left( \frac{\sb}{\Ub} \right) \, , 
\end{align}
which is much greater than $1/\sb$ (i.e., $\tc \sb \gg 1$) and is approximately independent of the population size \citep{desai:etal:2013}. Thus, like the rare driver regime above, the number of deleterious passengers does not necessarily decrease in larger and more rapidly adapting populations. Yet in this case, the cost of a typical passenger does not increase as rapidly with $N$ or $\Ub$, which reflects the fact that adaptation is not limited by the supply of beneficial mutations.  

Corrections to the beneficial substitution rate can be obtained in a similar way, although the algebra is more involved because $\etarb$, $g(x)$, and $h(x)$ are not independent. We carry out this calculation in \app{appendix:first-order}, and we find that the leading order correction to $\Rb$ is given by 
\begin{align}
\label{eq:full-etarb}
\begin{aligned}
\etarb & =  \frac{2 \vo}{(\xc-\sb)\sd} \left[ 1 - e^{-\frac{\xc \sd}{\vo}} \left[ \frac{\xc}{\sb} \left( e^{\frac{\sb \sd}{\vo}} - 1 \right) + 1 \right] \right] \\
	& \quad - \frac{\xc}{\sb} \left( e^{\frac{\sb \sd}{\vo}}-1 \right) e^{-\frac{\xc \sd}{\vo}} \, .
\end{aligned}
\end{align}
We compare this formula with simulations in \fig{fig:delta-collapse}. In sufficiently large populations where $\xc \gg \sb$, \eq{eq:full-etarb} reduces to the simple form
\begin{align}
\etarb \approx \frac{2 \vo}{\xc \sd} - 2 e^{-\frac{\xc \sd}{\vo}} \left[ \frac{\vo}{\xc \sd} + 1 + \frac{\xc \sd}{2 \vo} \right] \, ,
\end{align}
which depends only on the compound parameter $\frac{\xc \sd}{\vo} \approx \tc \sd$. In the limit that $\frac{\xc \sd}{\vo} \to 0$, $\etarb \approx 0 + \mathcal{O}\left(\frac{\xc \sd}{\vo}\right)^2$, which is consistent with the equal accumulation of nearly-neutral passengers in the nose and in the bulk population \citep{desai:fisher:2007}. On the other hand, when $\frac{\xc \sd}{\vo} \to \infty$, the reduction in $\Rb$ is consistent with a simple reduction in population size, $N_e = N(1-\Ud/\sd)$, similar to the ``ruby in the rough'' limit of the rare driver regime above. In between these two extremes, \eq{eq:full-etarb} predicts a maximum reduction in $\Rb$ at an intermediate value of $\frac{\xc \sd}{\vo} \approx 3.4$. This is consistent with our heuristic argument that passengers should only influence the fates of drivers if they are purged on the same timescale that determines the fate of a driver mutation, $\tb \sim \tsw$. Unlike the rare driver regime, the maximally interfering mutations in this case are approximately the same size as a typical passenger mutation ($\tb^{-1} \sim \tc^{-1}$), and are much smaller than the size of a typical driver ($\tb^{-1} \ll \sb$). 

\subsection*{Distributions of fitness effects}

\noindent Our analysis has so far assumed that the DFE is given by the simple two-effect form in \eq{eq:two-effect-model}. On the surface, such an assumption seems to conflict with empirical measurements of the DFE, which typically involve at least severalfold variation in the magnitudes of beneficial and deleterious mutations \citep{eyre-walker:keightley:2007}. In this section, we discuss how our analysis can be extended to this more biologically realistic scenario. 

At the level of approximation considered here, distributions of \emph{deleterious} fitness effects do not pose any major problems to our analysis. Since we have focused only on the leading order contributions in $\Ud$, the net effects of a deleterious DFE can be obtained simply by averaging over the deleterious effect sizes,
\begin{subequations}
\begin{align}
\etard[\rhod(s)] & = \int \left(\frac{\sd}{\sdbar}\right) \etard(\sd) \rhod(\sd) \, d\sd \, ,\label{eq:average-etard} \\
\etarb[\rhod(s)] & = \int \etarb(\sd) \rhod(\sd) \, d\sd \, ,
\end{align}
\end{subequations}
where $\etard(\sd)$ and $\etarb(\sd)$ are given by \eqs{eq:full-etard}{eq:full-etarb}. The outcome of this average can depend rather sensitively on both the supply of beneficial mutations [which controls the quantitative dependence of $\etard(\sd)$ and $\etarb(\sd)$] and the relative variation in $\rhod(\sd)$. In the broad distribution limit where $\rhod(\sd)$ is approximately uniform, the reduction in the beneficial substitution rate will be dominated by the density of deleterious mutations near $\sdb \sim \tb^{-1}$. The effects on the deleterious substitution rate are more subtle. When $\etard(\sd)$ decays exponentially with the fitness cost, $\Rd$ will be dominated by mutations near $\sdc \sim \tc^{-1}$ as expected. However, when $\etard(\sd)$ decays as an inverse power of $\sd$, the weighting factor in \eq{eq:average-etard} will remove much of the $\tc$ dependence in the integrand. In this case, significant contributions to $\Rd$ will arise from all deleterious mutations in the range $\tc^{-1} < \sd < \tb^{-1}$.   

\begin{figure}
\centering
\includegraphics[width=0.9\columnwidth]{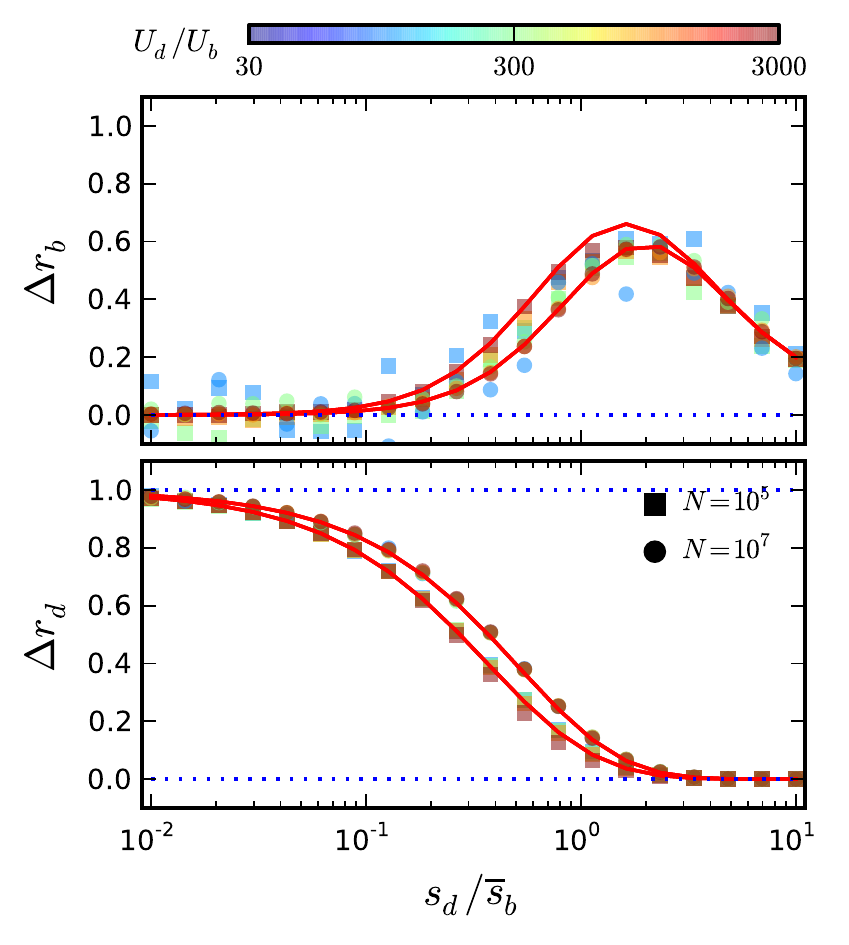}
\caption{Corrections to the substitution rate for an exponential distribution of beneficial fitness effects. Symbols denote the results of forward-time simulations for $N=10^5$ (squares) and $N=10^7$ (circles), with the remaining parameters the same as \fig{fig:delta-collapse}. Once again, $\etarb$ and $\etard$ are calculated using the simulated value of $\vo$, with $\xc$ estimated from \eq{eq:auxilliary-condition} and $\sb^*$ and $\Ub^*$ estimated from \eq{eq:exponential-effective-parameters}. For comparison, the solid red lines show the first-order predictions from \eq{eq:exponential-predictions}, using the same estimated values of $\vo$, $\xc$, and $\sb^*$. \label{fig:exponential-collapse}}
\end{figure}

In the rare driver regime, a similar averaging scheme can be applied for a distribution of \emph{beneficial} fitness effects, so that $\tc \approx 1/(2N\Ub\sbbar)$. However, this average breaks down in the multiple driver regime, where $\pfix(s)$ is not a simple linear function of $\Ub$. Instead, previous work has shown that a large class of beneficial DFEs can be approximated by a single-effect DFE, $\Ub\rhob(s) \approx \Ub^* \delta(s-\sb^*)$, provided that the effective selection coefficient and the mutation rate are chosen appropriately \citep{hegreness:etal:2006, desai:fisher:2007, good:etal:2012}. In agreement with this earlier work, we find that the single-$\sb$ approximation holds for $\etarb$ and $\etard$ as long as the effective parameters are associated with the expressions in \citet{good:etal:2012}: 
\begin{subequations} 
\label{eq:effective-parameters}
\begin{align}
\sb^* & = \xc + \vo \partial_s \log(\sb^*) \, , \\
\Ub^* & = \Ub \rhob(\sb^*) \sqrt{\frac{2\pi\vo}{1-\vo \partial_s^2 \rhob(\sb^*)}} \, .
\end{align}
\end{subequations}
Thus, we can extend our results to a distribution of fitness effects simply by replacing $\sb \to \sb^*$ and $\Ub \to \Ub^*$ in all of our previous expressions. The only difference is that $\sb^*$ and $\Ub^*$ now vary as a function of the population size and mutation rate, so the scaling of the various quantities (e.g., $\tc$ and $\tb$) can change.

As an example, we consider the case where beneficial mutations follow an exponential distribution, $\rhob(s) \propto \exp \left( -s/\sbbar \right)$. In this case, we have previously shown that the effective selection strength and mutation rate are given by 
\begin{align}
\label{eq:exponential-effective-parameters}
\sb^* = \xc - \frac{\vo}{\sbbar} \, , \quad \Ub^* = \Ub \sqrt{\frac{2 \pi \vo}{\sbbar^2}} e^{-\xc/\sbbar} \, ,
\end{align}
where $\xc \approx \sb^*$ for all but the largest population sizes \citep{good:etal:2012}. Thus, the fittest individuals in the population typically contain only one more driver mutation than the mean, and the dynamics of adaptation bear some resemblance to the selective sweeps picture above. In this ``quasi-sweep'' regime, the size of a typical driver increases with the population size in such a way that the nose-to-mean time,
\begin{align}
\tsw \sim \frac{\xc}{\vo} \sim \frac{2 \log^2( \sbbar / \Ub )}{\sbbar \log(N\Ub)} \, ,
\end{align}
remains a decreasing function of $N$. This is similar to the rare driver regime above, although the dependence on $N$ is much weaker in this case. Substituting our expressions for $\sb^*$ and $\Ub^*$ into \eqs{eq:full-etard}{eq:full-etarb}, we find that the substitution rates are given by
\begin{subequations}
\label{eq:exponential-predictions}
\begin{align}
\etard & \approx \left( \frac{ 1 - e^{-\frac{\xc \sd}{\vo}}}{\frac{\xc \sd}{\vo}} \right) \exp \left[ -\frac{\sd}{\sbbar}  \right] \label{eq:exponential-etard} \, , \\
\etarb & \approx \frac{2 \sbbar}{\sd} - 2 e^{-\frac{\sd}{\sbbar}} \left[ \frac{\sbbar}{\sd} + \left( \frac{\sbbar}{\sd} + \frac{1}{2} \right) \left( 1 - e^{-\frac{\xc \sd}{\vo}} \right) \right] \, . \label{eq:exponential-etarb}
\end{align}
\end{subequations}
We compare these predictions to simulations in \fig{fig:exponential-collapse}. Similar to the single-$\sb$ case, we again observe a transition from a regime of effective neutrality ($\Rd \approx \Ud$) to a regime effective lethality ($\Rd \approx 0$) at a characteristic effect size $\sdc \sim \vo/\xc$. This threshold is much smaller than the size of an average beneficial mutation ($\sbbar$) as well as the size of a typical driver ($\sb^* \gg \sbbar$). In this case there are no independent estimates of $\tc$, but these results suggest that $\tc \sim \vo/\xc$ holds for an exponential DFE as well. However, like the rare driver regime, the deleterious substitution rate in \eq{eq:exponential-etard} decays as a power law when $\sd \gg \sdc$, as opposed to the exponential dependence in \eq{eq:approx-etard}. Similarly, the reduction in the beneficial substitution rate in \eq{eq:exponential-etarb} is maximized when $\sd \approx 1.7 \sbbar \ll \sb^*$, where hitchhiking is already unlikely ($\Rd \approx 0$). This illustrates a subtle feature of the single-$s$ equivalence principle employed above. If we condition on the size of a typical driver ($\sb^*$), our results are insensitive to the shape of the DFE. However, if we condition on the size of an average \emph{potential} mutation ($\sbbar$), then the relevant deleterious mutations can depend rather sensitively on the shape of $\rhob(s)$. 

\section*{Discussion}

\noindent In any sufficiently complex organism, spontaneous mutations will have a broad range of effects on reproductive fitness. This leads to a natural question: which (if any) of these mutations will influence the evolutionary dynamics of the population? If certain mutations are more important than others, it is possible to focus only on a subset of potential mutations? In the general case, these questions can be difficult to answer. Adaptive changes account for just a small fraction of all possible mutations, but when they do arise, beneficial variants are rapidly amplified by selection and dramatically alter the evolution of the population. Deleterious mutations, in contrast, have a negligible impact individually, but their greater numbers can nevertheless lead to a large collective influence. Given this competition between the scales of mutation and selection, it is possible that beneficial and deleterious mutations \emph{both} play an important role in certain populations. 

Here, we have introduced a quantitative mathematical framework for characterizing the effects of deleterious passengers in rapidly evolving populations. By leveraging previous results in the absence of deleterious mutations, we derived simple formulae for the rates of sequence evolution when beneficial and deleterious mutations possess a broad range of fitness effects. 
These results provide important qualitative intuition about the effects of deleterious passengers, since they allow us to estimate \emph{which} deleterious mutations are most likely to hitchhike to fixation and which will hinder the fixation of the drivers. 

In the case of hitchhiking, we found that the maximum cost of a passenger is determined by the inverse of the coalescent timescale, $\sd^* \sim \tc^{-1}$, which reduces to the traditional ``drift-barrier'' ($\sdc \approx N^{-1}$) in the absence of other mutations. When drivers are more common, the location of this neutral threshold can grow to be much larger than the inverse population size, with a qualitatively different dependence on $N$. Thus, as observed in previous studies \citep{schiffels:etal:2011}, larger and more rapidly adapting populations will often accumulate a larger number of more strongly deleterious mutations, though the total \emph{fraction} of deleterious substitutions must still decline. This increased deleterious load can have important implications for the stability of rapidly adapting proteins [e.g., in influenza \citep{strelkowa:laessig:2012}], and is likewise relevant when inferring the prevalence of adaptive mutations from changes in dN/dS ratios \citep{ostrow:etal:2014}. It is important to note, however, that the deleterious load can only increase with $N$ in the presence of strongly beneficial driver mutations ($\tc \sb \gg 1$), and not as a general consequence of linkage. Indeed, we find that $\sdc$ decreases (weakly) with $N$ in ``infinitesimal'' models of linked selection ($\tc \sb \ll 1$)  \citep{tsimring:etal:1996, cohen:kessler:levine:2005, hallatschek:2011, neher:hallatschek:2013, neher:etal:2013, good:etal:2014} and in the presence of Muller's ratchet \citep{soderberg:berg:2007}.  

In addition to the size of a successful passenger, our framework also allows us to identify mutations that are most likely to hinder fixation of the drivers. We saw that this influence is maximized for deleterious mutations of an intermediate effect $\sdb \approx \tb^{-1}$, set by the stochastic phase of a successful driver mutation. When drivers are rare, this is simply the drift-time $\tb \sim \sb^{-1}$, and deleterious mutations above this threshold limit the rate of adaptation through a well-known reduction in effective population size \cite{charlesworth:1994, peck:1994, orr:2000, wilke:2004}. However, when driver mutations interfere with each other ($\Rb \tc \gg 1$), this stochastic phase becomes much longer than $\sb^{-1}$, since multiple beneficial mutations are required for fixation. Thus, the relevant deleterious fitness effects are usually much smaller than the size of a typical driver, which emphasizes the importance of the two-effect DFE that we used throughout our analysis. Previous work has often focused on a simpler ``single-effect'' DFE, where the fitness effects of beneficial and deleterious mutations are identical ($\sb=\sd$) \citep{woodcock:higgs:1996, rouzine:etal:2003, rouzine:etal:2008, goyal:etal:2012}. Our present results suggest that these models may \emph{underestimate} the importance of deleterious mutations, since they implicitly neglect mutations with the largest potential influence. 

These findings suggest that the cumulative influence of deleterious passengers depends rather sensitively on the \emph{distribution} of beneficial and deleterious fitness effects, in addition to the population size and mutation rate. This makes it difficult to estimate the relevance of deleterious passengers in practice, since these distributions are only known to a very rough degree of approximation \citep{eyre-walker:keightley:2007}. The notable exception is in experimental microbial populations, where high-throughput screens have enabled a much more detailed characterization of the fitness effects of new mutations \citep{elena:etal:1998, wloch:etal:2001, sanjuan:etal:2004, kassen:bataillon:2006, qian:etal:2012, frenkel:etal:2014}. As a concrete example, we have recently estimated $\rhob(s)$ for a strain of \emph{Saccharomyces cerevisiae} in rich media to be an exponential distribution with $\Ub=10^{-4}$ and $\sbbar=9 \times 10^{-3}$ \citep{frenkel:etal:2014}, which is also consistent with recent measurements of the yeast deletion collection \citep{qian:etal:2012}. If we assume that $\rhod(s)$ can also be estimated from the deletion collection data, and that $N \sim 10^5$, then our theory implies that deleterious mutations will only start to influence the rate of adaptation when $\Ud \sim 10^{-2}$, already an order of magnitude larger than the total per-genome mutation rate in yeast \citep{lynch:etal:2008}. Deleterious passengers are therefore unlikely to impede the rate of adaptation in these populations. Recent work has shown that $\sbbar$ may decrease by as much as a factor of 4 over $\sim 1000$ generations of evolution due to the effects of diminishing returns epistasis \citep{khan:etal:2011, chou:etal:2011, wiser:etal:2013, kryazhimskiy:etal:2014}. This only lowers our estimate of $\Ud^*$ by a factor of 2, which suggests that deleterious mutations are also unlikely to contribute to the long-term differences in evolvability of these strains in the absence of more complicated epistatic interactions. Of course, it is not surprising that deleterious mutations play a small role here, since these populations have been studied precisely because they repeatedly adapt to their laboratory environment. In more well-adapted populations, mutator strains have been observed to evolve towards lower mutation rates \cite{mcdonald:etal:2012, maharjan:etal:2013, wielgoss:etal:2013}, suggesting that the deleterious load may eventually become more of a burden. However, without a more detailed estimate of the DFE, this hypothesis is merely speculative. 

In the present article, we have considered only the most basic effects of deleterious passengers on the long-term patterns of sequence evolution, leaving many potential avenues for future work. Most notably, we have omitted any discussion about the patterns of sequence diversity within the population, which provide a snapshot of the selective forces operating in the recent past. Deleterious passengers are potentially even more relevant for this contemporary data, since mutations that are too deleterious to fix can still generate appreciable fitness diversity if they are sufficiently common \citep{haigh:1978}. Our perturbative corrections to the fitness distribution can potentially capture some of this deleterious diversity, but the effects on neutral polymorphism and the implications for recombining chromosomes will require additional analysis. Recent work by \citet{weissman:hallatschek:2014} could potentially be used to address these questions. 

A second limitation of our analysis is that it is fundamentally perturbative in nature. By focusing only on the leading-order corrections to the dynamics in powers of $\Ud/\xc$, our results are primarily applicable when the net effect of deleterious mutations is small (i.e., $\vo-v \ll \vo$). Thus, we have explicitly neglected cases where many deleterious mutations fix cooperatively due to Muller's ratchet \citep{soderberg:berg:2007, mcfarland:etal:2013}, or the long-term ``fixed-point'' where the accumulation of beneficial and deleterious mutations balances \citep{goyal:etal:2012}. In practice, our expressions are often quite accurate, even permitting estimates of the $v \approx 0$ fixed-point in certain cases. However, a more thorough characterization of this fixed point (and the shapes of $\rhob(s)$ and $\rhod(s)$ that are attained there) remains an important avenue for future work. 

\begin{acknowledgments}
We thank Daniel Balick, Ivana Cvijovi\'{c}, and Karri DiPetrillo for useful discussions. This work was supported in part by a National Science Foundation Graduate Research Fellowship, the James S. McDonnell Foundation, the Alfred P. Sloan Foundation, the Harvard Milton Fund, grant PHY 1313638 from the NSF, and grant GM104239 from the NIH. Simulations in this paper were performed on the Odyssey cluster supported by the Research Computing Group at Harvard University. 
\end{acknowledgments}


\onecolumngrid
\appendix

\section{Stochastic model of the fitness distribution and the mean-field approximation}
\label{appendix:stochastic-model}

\noindent As described in the text, the dynamics of our evolutionary model can be recast in terms of the population fitness distribution, $f(X,t)$, which tracks the fraction of individuals in each ``fitness class'' $X$. In the diffusion limit, these fitness classes obey the Langevin dynamics, 
\begin{align}
\label{eq:fitness-class-langevin}
\frac{\partial f(X)}{\partial t} = \underbrace{\left[ X - \overline{X}(t) \right] f(X)}_\text{selection} + \underbrace{U \int ds \, \rho(s) \left[ f(X-s) - f(X) \right]}_\text{mutation} + \underbrace{\int dX' \, \left[ \delta(X-X') - f(X) \right] \sqrt{\frac{f(X')}{N}} \eta(X')}_\text{genetic drift} \, ,
\end{align}
where $\overline{X}(t) = \int dX \, X f(X,t)$ is the mean fitness of the population and $\eta(X)$ is a Brownian noise term \citep{good:desai:2013}. \Eq{eq:fitness-class-langevin} represents a natural generalization of the single-locus diffusion equation for a genome with a large number of selected sites. To track the fate of a lineage founded by an individual with fitness $X_0$, we introduce the labeled fitness classes $g(X,t)$ and $f(X,t)$ corresponding to the focal lineage and the background population, respectively. These fitness classes obey a generalized version of \eq{eq:fitness-class-langevin},
\begin{subequations}
\label{eq:full-lineage-langevin}
\begin{align}
\begin{aligned}
\frac{\partial f(X)}{\partial t} = & \left[ X - \overline{X}(t) \right] f(X) + U \int ds \, \rho(s) \left[ f(X-s)-f(X) \right] + \sqrt{\frac{f(X)}{N}} \eta_f(X) \\
	& - f(X) \left[ \int \sqrt{\frac{f(X')}{N}} \eta_f(X') + \sqrt{\frac{g(X')}{N}} \eta_g(X') \, dX' \right] \, , \end{aligned}  \\
\begin{aligned}
\frac{\partial g(X)}{\partial t} = & \left[ X - \overline{X}(t) \right] g(X) + U \int ds \, \rho(s) \left[ g(X-s)-g(X) \right] +  \sqrt{\frac{g(X)}{N}} \eta_g(X) \\
	&  - g(X) \left[ \int \sqrt{\frac{f(X')}{N}} \eta_f(X') + \sqrt{\frac{g(X')}{N}} \eta_g(X') \, dX' \right] \, ,
\end{aligned}
\end{align}
\end{subequations}
with the initial condition $g(X,0) = \frac{1}{N} \delta(X-X_0)$. For general $N$ and $U\rho(s)$, there is no closed form solution of \eq{eq:lineage-langevin}. However, in sufficiently large populations we can employ a ``mean-field'' approximation that has been used in several previous studies \citep{neher:etal:2010, neher:shraiman:draft:2011, good:etal:2012, fisher:2013}. 

The basic idea behind this approximation is that there is a separation of frequency scales between the regime where genetic drift is important and the regime where population saturation is important \citep{desai:fisher:2007}. In other words, the fate of a lineage is determined while it is still rare ($\int g(X) \, dX \ll 1$), while most of the remaining population evolves deterministically. From these assumptions, we can rewrite \eq{eq:lineage-langevin} in the simpler form
\begin{subequations}
\label{eq:lineage-langevin}
\begin{align}
\frac{\partial f(X)}{\partial t} & = \left[ X - \int X' f(X') \, dX'  \right] f(X) + U \int ds \, \rho(s) \left[ f(X-s)-f(X) \right] \label{eq:original-f-equation} \, , \\
\frac{\partial g(X)}{\partial t} & = \left[ X - \int X' f(X') \, dX' \right] g(X) + U \int ds \, \rho(s) \left[ g(X-s)-g(X) \right] + \sqrt{\frac{g(X)}{N}} \eta_g(X) \, ,
\end{align}
\end{subequations}
where we have approximated the dynamics of the focal lineage by a simple linear branching process. The fixation probability of this lineage can then be deduced using standard techniques from the theory of branching processes. We introduce the generating functional $H[\phi(X),t] = \left\langle \exp \left[ - \int N \phi(X) g(X,t) \, dX \right] \right\rangle$, which satisfies the partial differential equation
\begin{align}
\label{eq:generating-functional}
\frac{\partial H}{\partial t} = \int \left\{ \left[ X - \int X' f(X',t) dX' \right] \phi(X) + U \int ds \, \rho(s) \left[ \phi(X+s) - \phi(X) \right] - \frac{\phi(x)^2}{2} \right\} \frac{\partial H}{\partial \phi(X)} \, dX \, ,
\end{align} 
with the initial condition $H[\phi(X),0] \approx 1-\phi \left( X_0 \right) + \mathcal{O}(N^{-1})$. Since we neglect saturation effects, we know that at long times the focal lineage must either go extinct or diverge to infinity. This yields a relation between the generating function and the fixation probability, 
\begin{align}
\lim_{t \to \infty} H[\phi(X),t] = 1- \mathrm{Pr}\left[ \int g(X) \theta[\phi(X)] \, dX > 0 \right] \, ,
\end{align} 
where $\theta(x)$ is the Heaviside step function. Thus, we must simply solve \eq{eq:generating-functional} to obtain the fixation probability for any subset of the focal lineage. We can achieve this via the method of characteristics. Letting $\tau$ denote backwards time, the characteristic equation for $\phi(X)$ is given by 
\begin{align}
\label{eq:phi-equation}
- \frac{\partial \phi(X)}{\partial \tau} & = \left[ X-\int X' f(X',t-\tau) \, dX' \right] \phi(X) + U \int ds \, \rho(s) \left[ \phi(X-s) - \phi(X) \right] - \frac{\phi(X)^2}{2} \, ,
\end{align}
with the backwards-time initial condition $\phi(X,\tau=0) = \phi(X)$. Meanwhile, the characteristic for $H$ is simply $\partial_\tau H = 0$, so we can immediately conclude that $H[\phi(X),t] = 1-\phi(X_0,\tau=t)$, and therefore that the fixation probability for a subset of the focal lineage is given by
\begin{align}
\label{eq:phi-fixation}
\mathrm{Pr}\left[ \int g(X) \theta[\phi(X,\tau=0)] \, dX > 0 \right] = \lim_{t \to \infty} \phi(X_0,\tau=t) \, . 
\end{align}
Finally, we let $w(X)$ be the unique long-time limit of $\phi(X,\tau)$ when $\phi(X) > 0$, or
\begin{align}
\label{eq:w-fixation}
w(X) \equiv \mathrm{Pr}\left[ \int g(X') \, dX' > 0 \right] = \lim_{t \to \infty} \phi(X,\tau=t) \, .
\end{align}
In other words, $w(X)$ is simply the fixation probability of a new lineage founded by a single individual with fitness $X$ at time $t=0$. Similar derivations of \eq{eq:w-fixation} can be found in \citet{good:etal:2012} and \citet{fisher:2013}, although we will sometimes require the more general expression in \eq{eq:phi-fixation}. 

Within our mean-field approximation, the fixation probability, $\pfix(s)$, for a new mutation with fitness effect $s$ can be obtained by averaging over the fitness backgrounds that the mutation could have arisen on. By construction, this distribution of backgrounds is simply $f(X)$, so that
\begin{align}
\label{eq:original-fixation-probability}
\pfix(s) = \int f(X-s) w(X) \, dX \, .
\end{align}
We must then match the ``microscopic'' dynamics of $\pfix(s)$ with the deterministic solution for $f(X,t)$ to obtain a self-consistent description of the evolutionary dynamics \citep{neher:etal:2010, neher:shraiman:draft:2011, hallatschek:2011, good:etal:2012, fisher:2013}. We carry out this procedure for several simple cases below. 

\section{Zeroth-order solution without deleterious mutations}
\label{appendix:zeroth-order}

\noindent In this section, we review the solution of the mean-field model in the absence of deleterious mutations ($\Ud=0$). This ``zeroth-order'' solution will form the basis of the perturbative analysis described in the text. Mirroring our discussion in the text, we distinguish between the rare driver regime and the multiple driver regime. 

\subsection{Rare driver mutations}

\noindent When driver mutations are rare [also known as the \emph{successional mutations} or \emph{strong-selection weak mutation (SSWM)} regime \citep{desai:fisher:2007}], multiple beneficial mutations rarely segregate in the population at the same time. We can therefore neglect the mutation terms in the mean-field approximation in \eqs{eq:original-f-equation}{eq:phi-equation}. When a beneficial mutation arises, the population is fixed for a single genotype, which (without loss of generality) can be assumed to have fitness $X=0$. In other words, the deterministic solution for the fitness distribution is simply $f(X,t) \approx \delta(X)$. The differential equation for $\phi(X,\tau)$ in \eq{eq:phi-equation} reduces to
\begin{align}
-\partial_\tau \phi \approx X \phi(X) - \frac{\phi(X)^2}{2} \, ,
\end{align}
from which we can obtain the long-term solution for the fixation probability,
\begin{align}
w(X) = \lim_{\tau \to \infty} \phi(X,\tau) = \begin{cases}
2X & \text{if $X > 0$,} \\
0 & \text{if $X < 0$.}
\end{cases}
\end{align}
Given $w(X)$, the fixation probability of a new mutation trivially follows from \eq{eq:original-fixation-probability},
\begin{align}
\pfix(s) \approx \begin{cases}
2s & \text{if $s > 0$,} \\
0 & \text{if $s=0$,}
\end{cases}
\end{align}
which we immediately recognize as the large-$N$ limit of the single-locus fixation probability, $\pfix(s) = 2s/(1-e^{-2Ns})$. The beneficial substitution rate is therefore given by $\Rb = 2N\Ub\sb$. In order for this solution to be self-consistent, we require that no new beneficial mutations establish during the fixation time of the driver, $\tfix = \frac{2}{\sb} \log ( 2 N\sb )$, or 
\begin{align}
\label{eq:clonal-interference-condition}
\Rb \cdot \tfix = 4N\Ub \log(2N\sb) \ll 1 \, .
\end{align}
This also implies that $\Rb \ll \sb$. 

\subsection{Multiple driver mutations}

\noindent In large populations, the rare-driver condition in \eq{eq:clonal-interference-condition} breaks down and multiple beneficial mutations will segregate in the population at the same time. In this case, the mean-field fitness distribution is no longer a $\delta$-function (with discrete jumps when drivers fix), but rather an extended traveling wave $f(X,t) = f(X-vt)$ that steadily increases in fitness at rate $v$. Without loss of generality, we can assume that the mean fitness of the population is zero when the new mutation arises. Then we can change variables to the \emph{relative} fitness, $x \equiv X - \overline{X}(t) \approx X-vt$, so that \eqs{eq:original-f-equation}{eq:phi-equation} become  
\begin{subequations}
\label{eq:evolutionary-equations}
\begin{align}
- v \partial_x f(x) & = x f(x) + \Ub \int ds \, \rhob(s) \left[ f(x-s) - f(x) \right] \, , \label{eq:appendix-f-equation} \\
v \partial_x w(x) & = x w(x) + \Ub \int ds \, \rhob(s) \left[ w(x+s) - w(x) \right] - \frac{w(x)^2}{2} \label{eq:appendix-w-equation} \, ,
\end{align} 
\end{subequations}
with $\pfix(s) = \int f(x-s) w(x) \, dx$, as described in the text. For this solution to be self-consistent, we require that $\pfix(0) \approx 1/N$, i.e., exactly one individual from the current population will become a future common ancestor \citep{hallatschek:2011, good:etal:2012, fisher:2013}. This serves to completely determine $v$, $f(x)$ and $w(x)$ as a function of $N$ and $\Ub \rhob(s)$. There are many different regimes to consider [see \citet{fisher:2013} for additional discussion], but we will focus on a particular case that is relevant for many microbial evolution experiments. We assume that the typical background fitness of a successful driver is much larger than the standard deviation of the fitness distribution ($x_{bg} \gg \sigma$). This ensures that there is a substantial amount of clonal interference in the population, which is consistent with the empirical observation that the fates of drivers strongly depend on background fitness \citep{lang:etal:2013}. Second, we assume that the fitness effect of a typical driver is also much larger than the standard deviation of the fitness distribution ($\sb^* \gg \sigma$). This ensures that there is not too much clonal interference in the population, and is consistent with direct measurements of the fitness distribution \citep{desai:etal:2007} and forward-time simulations using parameters inferred from marker divergence experiments \citep{frenkel:etal:2014}. In terms of the underlying parameters, a necessary (but not sufficient) condition for $\sb^* \gg \sigma$ is that selection is much stronger than mutation ($\sb^* \gg \Ub^*$). We consider the opposite regime in the following section. 

Assuming that these two conditions are met, we can use the approximate solution to \eq{eq:evolutionary-equations} derived in \citet{good:etal:2012}. This earlier work focused on the rate of adaptation and the distribution of fixed beneficial mutations, which were relatively insensitive to the approximate forms of $f(x)$ and $w(x)$ that we employed. In contrast, the perturbative analysis described in the text is much more sensitive to the precise details of our approximation scheme. Fortunately, we can obtain a suitable generalization of the analysis in \citet{good:etal:2012} by enforcing a basic symmetry constraint: we demand that the approximate expression for $w(x)$ varies self-consistently under infinitesimal boosts (i.e., changes in $v$) and translations (changes in $\overline{X}$). The resulting solution will still be incorrect in several key ways [see \citet{fisher:2013} for additional discussion], but it will be sufficient to calculate the leading-order corrections from deleterious mutations at the level of approximation required here.   

When $\sb^* \gg \sigma$, the contributions from the mutation terms in \eq{eq:evolutionary-equations} are small for most of the relevant fitnesses. For sufficiently large $x$, the fixation probability satisfies the reduced equation
\begin{align}
\label{eq:shoulder-equation}
v \partial_x w(x) \approx x w(x) - \frac{w(x)^2}{2} \, ,
\end{align}
which has the solution
\begin{align}
\label{eq:shoulder-w}
w(x) = \frac{2\xc e^{\frac{x^2-\xc^2}{2v}}}{1+\left( \frac{\xc}{x} \right) e^{\frac{x^2-\xc^2}{2v}}} \approx \begin{cases}
2x & \text{if $x > \xc$,} \\
2\xc e^{\frac{x^2-\xc^2}{2v}} & \text{if $x < \xc$.} 
\end{cases}
\end{align}
Here, $\xc \gg \sqrt{v}$ is a constant of integration that must be set to ensure that $w(x)$ matches on to the correct branch of the solution for smaller $x$. The solution in \eq{eq:shoulder-w} has a characteristic ``shoulder'' shape. For $x-\xc \gg \frac{v}{\xc}$, the fixation probability saturates to the Haldane limit $w(x) \approx 2x$, which reflects a dominant balance between the selection [$x w(x)$] and drift [$w(x)^2/2$] terms in \eq{eq:shoulder-equation}. In this regime, a new mutation will fix provided that it survives genetic drift \citep{good:etal:2012}. For $\xc-x \gg \frac{v}{\xc}$, the fixation probability is rapidly reduced due to clonal interference, which reflects a dominant balance between the selection [$xw(x)$] and mean fitness [$v\partial_x w(x)$] terms in \eq{eq:shoulder-equation}. In this regime, a new mutation will fix only if it can generate further beneficial mutations to outrun the steady increase in the mean fitness. The width of the crossover between these two regimes is of order $v/\xc$,  which becomes increasingly sharp in the limit that $\xc \gg \sqrt{v}$. 

Since the fixation probability must vanish when $x \to -\infty$, it is clear that the shoulder solution breaks down for smaller values of $x$ where the effects of ``lucky'' beneficial mutations become important. In this regime, $w(x)$ will depend rather sensitively on the precise shape of the DFE, since a mutation will only survive if it is rescued by an usually large driver mutation. Yet by definition, \emph{successful} mutations that land in this regime are atypical, and represent a small fraction of all substitutions. We therefore introduce a negligible amount of error by assuming that $w(x)$ vanishes below a certain threshold, so that
\begin{align}
\label{eq:appendix-approx-w}
w(x) & = \begin{cases}
2x & \text{if $x > \xc$,} \\
2x_c e^{\frac{x^2-\xc^2}{2v_0}} & \text{if $\xmin < x < \xc$,} \\
0 & \text{else.}
\end{cases}
\end{align}  
Note that the functional form of \eq{eq:approx-w} is independent of $\Ub \rhob(s)$, which only enters through location of $\xmin$ and $\xc$. In \citet{good:etal:2012}, we had previously assumed that $\xmin \approx 0$, but if taken literally this leads to the pathological behavior pointed out by \citet{fisher:2013}. For the present analysis, we will only assume that $\xc - \xmin \lesssim \sb^*$, which is sufficient to ensure that $\xmin$ drops out of the analysis in Good \emph{et al} (2012). The precise value of $\xmin$ will be set by the symmetry considerations below. As described in \citet{good:etal:2012}, the location of $\xc$ can be obtained from an integral transform of \eq{eq:appendix-w-equation}, which reduces to
\begin{align}
\Ub \int dx \, \int ds \, \rhob(s) e^{-\frac{(x-s_b)^2}{2 \vo}} w(x) \approx \int dx \, e^{-\frac{x^2}{2 \vo}} \frac{w(x)^2}{2} \nonumber
\end{align}
in the limit that $\xc-\sb^* \gg \sqrt{v}$ and $\sb^* \gg \sqrt{v}$. In order to satisfy the symmetry constraints below, we will need to introduce an $\mathcal{O}(1)$ ``fudge factor'' $\F$, so that this relation is instead given by 
\begin{align}
\Ub \int dx \, \int ds \, \rhob(s) e^{-\frac{(x-s_b)^2}{2 \vo}} U_b w(x) \approx \F \int dx \, e^{-\frac{x^2}{2 \vo}} \frac{w(x)^2}{2} \, .
\end{align}
Then we can substitute our approximate expression for $w(x)$ from \eq{eq:appendix-approx-w} and evaluate the integrals to obtain a condition for $\xc$,              
\begin{align}
1 & = \frac{\Ub^*}{2 \F \sb^*} \left[ 1 - \frac{\sb^*}{\xc} \right]^{-1} e^{\frac{\xc \sb^*}{v}-\frac{{\sb^*}^2}{2 v}} \,  ,
\end{align}
where the effective selection coefficient and mutation rate are given by \eq{eq:effective-parameters} in the main text. Together, these expressions completely determine $w(x)$ as a function of $v$ and $\Ub\rhob(s)$. 

An approximate expression for the fitness distribution can be obtained in a similar manner. Ignoring the mutation term, $f(x)$ satisfies the reduced equation $-\partial_x f(x) = x f(x)$, which has a simple Gaussian solution with mean zero and variance $\sigma^2 = v$. This solution is valid for fitnesses up to $x \approx \xc$, after which the input from the mutation term causes $f(x)$ to rapidly approach zero and eventually turn negative \citep{rouzine:etal:2008, goyal:etal:2012, fisher:2013}. This is an artifact of our mean-field approximation, which neglects the increasingly important effects of drift near $x\approx\xc$. However, like the behavior of the fixation probability for $x \ll \xmin$, mutations that originate from $x \gg \xc$ are highly atypical (since they rely on a chance fluctuation of the fitness distribution), and therefore constitute a relatively small fraction of all substitutions. We therefore introduce a negligible amount of error by assuming that $f(x)$ vanishes above $\xc$, so that
\begin{align}
f(x) \approx \begin{cases}
0 & \text{if $x > \xc$,} \\
\frac{1}{\sqrt{2\pi v}} e^{\frac{-x^2}{2\pi v}} & \text{if $x<\xc$.}
\end{cases}
\end{align}
Then we can substitute our approximate expressions for $f(x)$ and $w(x)$ into the consistency condition $\pi(0) \equiv \int f(x) w(x) \, dx \approx \frac{1}{N}$, and obtain a second relation
\begin{align}
\label{eq:population-condition}
1 & = \frac{2 N \xc(\xc-\xmin)}{\sqrt{2 \pi v}} e^{-\frac{\xc^2}{2v}} \, ,
\end{align}
which uniquely determines $v$ as a function of $N$ and $U_b \rho(s)$.

To determine $\xmin$ and $\F$, we will enforce a symmetry constraint on our approximate solution for $w(x)$. Specifically, we will assume that under infinitesimal boosts and translations, we should get the same expression for $w(x)$ whether we expand our approximate solution or solve \eq{eq:appendix-w-equation} perturbatively. Although this may seem like an abstract requirement, these symmetry transformations turn out to be intimately related to the $\sd \to 0$ and $\sd \to \infty$ limits of the full deleterious model. 

First, suppose that we perform an infinitesimal boost by perturbing $v = \vo(1+\epsilon)$, with $\epsilon \ll 1$. This is equivalent to adding an additional term $\epsilon \vo \partial_x w(x)$ to the left hand side of \eq{eq:appendix-w-equation}. According to our analysis above, the non-perturbative solution is 
\begin{align}
w(x) = \begin{cases}
    2x & \text{if $x > \xc(1+\delta)$,} \\
    2\xc(1+\delta_x) e^{\frac{x^2-\xc^2 (1+\delta_x)^2}{2 \vo (1+\epsilon)}} & \text{if $x < \xc(1+\delta_x)$,} \\
\end{cases}    
\end{align}
where $\delta_x$ and (and the corresponding $\delta_s$ and $\delta_U$) are defined by
\begin{subequations}
\begin{align}
& 1 = \frac{\Ub^*(1+\delta_U)}{2 \F \sb^*(1+\delta_s)} \left[ 1 - \frac{\sb^*(1+\delta_s)}{\xc(1+\delta_x)} \right]^{-1} e^{\frac{\xc \sb^*(1+\delta_x)(1+\delta_s)}{\vo (1+\epsilon)}-\frac{{\sb^*}^2(1+\delta_s)^2}{2 \vo(1+\epsilon)}} \, , \\
& \sb^*(1+\delta_s) = \xc(1+\delta_x) + \vo(1+\epsilon) \partial_s \log \rhob\left[ \sb^*(1+\delta_s) \right] \, , \\
& \Ub^*(1+\delta_U) = U_b \rhob\left[ \sb^*(1+\delta_s) \right] \sqrt{\frac{2 \pi \vo(1+\epsilon)}{1-\vo(1+\epsilon) \partial_s^2  \rhob\left[ \sb^*(1+\delta_s) \right] }} \, ,
\end{align}
\end{subequations}
From the definition of $\sb^*$, there are no contributions to $\delta_x$ from $\delta_s$ and $\delta_U$ to lowest order in $\epsilon$, which shows that 
\begin{align}
\delta_x 
\approx \left(1-\frac{\sb^*}{2\xc} \right) \epsilon \, , 
\end{align}
and
\begin{align}
\label{eq:semiperturbative-w}
w(x) \approx \begin{cases}
\wo(x) & \text{if $x > \xc \left[ 1 + \left( 1 - \frac{\sb^*}{2\xc} \right) \epsilon \right]$,} \\
\xc e^{\frac{x^2-\xc^2}{2\vo}} + \mathcal{O}(\epsilon) & \text{if $\xc < x < \xc \left[ 1 + \left( 1 - \frac{\sb^*}{2\xc} \right) \epsilon \right]$,} \\
\wo(x) \left[ 1 + \epsilon g(x) \right]  & \text{if $x < \xc$,}
\end{cases}
\end{align}
where we have defined $g(x) = -\frac{x^2}{2 \vo} - \frac{\xc^2}{2\vo} \left( 1 - \frac{\sb^*}{\xc} \right)$. Then for any test function $\zeta(x)$, we have
\begin{align}
\int \zeta(x) w(x) \, dx & = \int \zeta(x) \wo(x) \, dx + \epsilon \int_{-\infty}^{\xc} \zeta(x) g(x) \nonumber \\
	& \quad + \int_{\xc}^{\xc + \epsilon \xc \left( 1 - \frac{\sb^*}{2\xc} \right)} \eta(x) \left[ 2\xc e^{\frac{x^2-\xc^2}{2\vo}} - 2x + \mathcal{O}(\epsilon) \right] \, dx \, , \nonumber \\
	& = \int \zeta(x) \wo(x) \, dx + \epsilon \int_{-\infty}^{\xc} \zeta(x) g(x) + \mathcal{O}(\epsilon^2) \, ,
\end{align}
and we see that the contribution from the nonperturbative boundary layer between $\xc$ and $\xc + \epsilon \xc \left( 1 - \frac{\sb^*}{2\xc} \right)$ vanishes to lowest order in $\epsilon$. Thus, by solving \eq{eq:appendix-w-equation} and then expanding, we find that
\begin{align}
w(x) \approx \begin{cases}
\label{eq:perturbative-w}
\wo(x) \left[ 1 - \epsilon \cdot 0 \right] & \text{if $x > \xc$,} \\
\wo(x) \left[ 1 - \epsilon \left( \frac{x^2}{2\vo} + \frac{\xc^2}{2\vo} \left( 1 - \frac{\sb^*}{\xc} \right) \right) \right] & \text{if $x < \xc$.}
\end{cases}
\end{align}
Now we try to obtain this solution by expanding \eq{eq:appendix-w-equation} in powers of $\epsilon$ and solving perturbatively. To that end, we rewrite $w(x)$ in the form $w(x) = \wo(x) \left[ 1 + \epsilon g(x) \right]$, which yields a related equation for $g(x)$:
\begin{align}
\label{eq:boost-g}
\partial_x g(x) = - \partial_x \log \wo(x) - \frac{\wo(x) g(x)}{2\vo} + \frac{U_b}{\vo} \left[ \frac{\wo(x+s) g(x+s)}{\wo(x)} - g(x) \right]   \, ,
\end{align}
We can solve this equation using the same approximation methods that we used for \eq{eq:appendix-w-equation} above. For $x \gg \xc$, the mutation and mean fitness terms can be ignored, and we have
\begin{align}
g(x) = - \frac{\vo}{x^2} \approx 0 \, .
\end{align}
Meanwhile, for $x < \xc$, the drift and mutation terms can be ignored, and we find that
\begin{align}
g(x) = - \frac{x^2}{2\vo} + C \, ,
\end{align}
where $C$ is a constant of integration. Since $w(x)$ is continuous at $x=\xc$, it is tempting to fix $C$ by demanding that $g(x)$ is also continuous at $\xc$, but this would not be correct. We saw in \eq{eq:semiperturbative-w} that $w(x)$ develops a nonperturbative boundary layer between $\xc$ and $\xc + \epsilon \xc ( 1 - \sb^* / 2 \xc )$. In this region (which does not contribute to any integrals at leading order), $w(x)$ changes rapidly to ensure continuity, but the perturbative correction $g(x)$ is \emph{not} continuous. Instead, we will fix $C$ using the integral transform above that we used to determine $\xc$. This yields the relation
\begin{align}
\int^{\xc} e^{-\frac{(x-\sb)^2}{2\vo}} \wo(x) g(x) \, dx - \int^{\xc} e^{-\frac{x^2}{2\vo}} \vo \partial_x \wo(x) \, dx = \F \int^{\xc} e^{-\frac{x^2}{2 \vo}} \wo(x)^2 g(x) \, dx \, ,
\end{align}  
which reduces to 
\begin{align}
g(\xc) \int^{\xc} e^{-\frac{(x-s_b)^2}{2 \vo}} \wo(x) \, dx - \int^{\xc}_{\xmin} x \xc e^{-\frac{\xc^2}{2\vo}}\, dx = 2 g(\xc) \F \int^{\xc} e^{-\frac{x^2}{2 \vo}} \wo(x)^2 \, dx \, ,
\end{align}  
and hence
\begin{align}
C = \frac{\xc^2}{2\vo} - \frac{\xc^2}{\vo} \left( \frac{\xc - \xmin}{s_b} \right)\frac{2 - \frac{\xc-\xmin}{\xc}}{4 \F} \, .
\end{align}
Thus, we see that the perturbative solution matches our previous expression in \eq{eq:perturbative-w} provided that 
\begin{align}
\label{eq:boost-relation}
\left( \frac{\xc-\xmin}{\sb^*} \right)^2 - 2 \left( \frac{\xc}{\sb^*} \right) \left( \frac{\xc-\xmin}{\sb^*} \right) + 2 F \left( \frac{2 \xc}{\sb^*} - 1 \right) = 0 \, .
\end{align}
This provides one relation between $\xmin$ and $F$, and is sufficient to show that $\xc-\xmin \sim \mathcal{O}(\sb^*)$.   

Now we repeat this analysis for an infinitesimal translation, $x \to x - \epsilon \xc$, which is equivalent to adding a $-\epsilon \xc w(x)$ term in \eq{eq:appendix-w-equation}. In this case, the nonperturbative solution is
\begin{align}
w(x) & = \begin{cases}
x-\epsilon \xc & \text{if $x > \xc + \epsilon \xc$,} \\
\xc e^{\frac{(x-\epsilon \xc)^2}{2\vo} - \frac{\xc^2}{2\vo}} & \text{if  $x < \xc + \epsilon \xc$,}
\end{cases}
\end{align}
which, when expanded to first order in $\epsilon$, yields
\begin{align}
w(x) & \approx \begin{cases}
\wo(x) \left[ 1 - \epsilon \frac{\xc}{x} \right] & \text{if $x > \xc + \epsilon xc$,} \\
\xc e^{\frac{x^2-\xc^2}{2 \vo}} + \mathcal{O}(\epsilon) & \text{if $\xc < x < \xc + \epsilon \xc$,} \\
\wo(x) \left[ 1 - \epsilon \frac{x \xc}{\vo} \right] & \text{if $x < \xc$.}
\end{cases}
\end{align}
Again, the boundary layer does not contribute to any integrals of $w(x)$ at leading order, so we can simply drop it from our expansion to obtain
\begin{align}
w(x) \approx \begin{cases}
\label{eq:perturbative-translation-w}
\wo(x) \left[ 1 - \epsilon \left( \frac{\xc}{x} \right) \right] & \text{if $x > \xc$,} \\
\wo(x) \left[ 1 - \epsilon \left( \frac{x \xc}{\vo} \right) \right] & \text{if $x < \xc$.}
\end{cases}
\end{align}
We now try to obtain this solution in the opposite order by expanding \eq{eq:appendix-w-equation} and solving perturbatively. Defining $w(x) = \wo(x) \left[ 1 + \epsilon g(x) \right]$, we see that $g(x)$ satisfies
\begin{align}
\partial_x g(x) & = - \frac{\xc}{\vo} - \frac{\wo(x) g(x)}{\vo} + \frac{U_b}{\vo} \left[ \frac{\wo(x+s) g(x+s)}{\wo(x)} - g(x) \right] \, ,    
\end{align}
whose solution is simply
\begin{align}
g(x) & = \begin{cases}
- \frac{\xc}{x} & \text{if $x > \xc$,} \\
-\frac{x \xc}{\vo} + C & \text{if $x < \xc$.}
\end{cases}
\end{align}
The constant $C$ is again determined by the integral condition 
\begin{align}
\int e^{-\frac{(x-s_b)^2}{2\vo}} \wo(x) g(x) \, dx - \xc \int e^{-\frac{x^2}{2\vo}} \wo(x) \, dx = \F \int e^{-\frac{x^2}{2 \vo}} \wo(x)^2 g(x) \, dx \, ,
\end{align}  
which shows that
\begin{align}
C = \frac{\xc^2}{\vo} - \frac{\xc^2}{\vo} \left[ \frac{\xc - \xmin }{2 \F s_b} \right] \, .
\end{align}
This perturbative solution matches our previous expression in \eq{eq:perturbative-translation-w} if
\begin{align}
\left(\frac{\xc-\xmin}{\sb} \right) = 2F \, , 
\end{align}
which provides a second relation between $\xmin$ and $F$. Combined with the first relation in \eq{eq:boost-relation}, this shows that
\begin{align}
\xmin = \xc-\sb^* \, , \quad F=\frac{1}{2} \, .
\end{align}
This completes our analysis of the zeroth-order solution in the multiple mutations regime. 

\subsection{The infinitesimal limit}

\noindent In addition to the ``strong driver'' limit ($\sb^* \gg \sqrt{v}$), we can also analyze the behavior of the multiple driver regime in the ``weak driver'' limit ($\sb^* \ll \sqrt{v}$). A particularly simple case is the \emph{infinitesimal limit}, where $\Ub \to \infty$ and $\sb \to 0$ in such a way that the product $2D \equiv \Ub \sb^2$ remains fixed \citep{tsimring:etal:1996, cohen:kessler:levine:2005, neher:etal:2010, neher:shraiman:draft:2011, hallatschek:2011, neher:hallatschek:2013, neher:etal:2013, good:etal:2014}. Since $\sb \to 0$, the driver mutations behave as if they were effectively neutral, so the substitution rate is simply
\begin{align}
\Rb \approx \Ub \, .
\end{align}
However, the overall evolutionary dynamics is far from neutral. Since we are simultaneously taking $\Ub \to \infty$, there are enough infinitesimal drivers segregating within the population that the total variance in fitness remains finite. It is straightforward to solve for the distribution of fitnesses and the fixation probability in this limit, since the mutation terms in \eq{eq:evolutionary-equations} can be expanded in a Taylor series, 
\begin{subequations}
\begin{align}
- \sigma^2 \partial_x f(x) & = x f(x) + D \partial_x^2 f(x) \, , \\
\sigma^2 \partial_x w(x) & = x w(x) + D \partial_x^2 w(x) - \frac{w(x)^2}{2} \, ,
\end{align}
\end{subequations}
where $\sigma^2 = v-Us$ is the variance in fitness within the population. In large populations, $w(x)$ again develops a sharp boundary layer near $x \approx \xc$, above which it approaches the Haldane form $w(x) \approx 2x$. Below $\xc$, $f(x)$ and $w(x)$ both satisfy a modified Airy equation, so that
\begin{align}
f(x) & \propto e^{-\frac{\sigma^2 x}{2D}} \mathrm{Ai}\left[ \frac{\sigma^4}{4 D^{4/3}} - \frac{x}{D^{1/3}} \right] \, , \\
w(x) & \propto e^{\frac{\sigma^2 x}{2 D}} \mathrm{Ai}\left[ \frac{\sigma^4}{4 D^{4/3}} - \frac{x}{D^{1/3}} \right] \, ,
\end{align}
where $\mathrm{Ai}(z)$ is the solution to the Airy equation that converges for large $z$ \citep{hallatschek:2011}. The full solution is obtained by matching $w(x)$ and its derivative at $x=\xc$. For large $N$, the argument of the Airy function will be close to the first zero, $z_0 \approx -2.33$ \citep{fisher:2013}. Expanding around this point, we find that 
\begin{align}
w(x) \approx \begin{cases}
2x & \text{if $x > \xc$,} \\
2\xc \left( \frac{\sigma^2}{2 D^{2/3}} \right) e^{\frac{\sigma^2}{2D} \left( x - \xc \right)} \mathrm{Ai} \left( \frac{\xc-x}{D^{1/3}} + z_0 + \frac{2 D^{2/3}}{\sigma^2} \right) & \text{else,}
\end{cases}
\end{align}
where the relationship between $\xc$ and $\sigma^2$ [i.e., the analogue of \eq{eq:auxilliary-condition}] is given by 
\begin{align}
\xc \approx \frac{\sigma^4}{4 D} \, .
\end{align}
Finally, we can solve for $\sigma^2$ by substituting these expressions into the self-consistency condition $\pi(0) \approx 1/N$, which yields
\begin{align}
\sigma^2 \approx \left[ 24 D^2 \log \left( 2 N D^{1/3} \right) \right]^{1/3}
\end{align}
in the limit of large $N$ \citep{tsimring:etal:1996, cohen:kessler:levine:2005, hallatschek:2011}. 

\section{Leading-order corrections from deleterious mutations}
\label{appendix:first-order}

\noindent In this section, we derive the leading order corrections to $\Rb$ and $\Rd$ in the presence of deleterious mutations. As with any perturbative calculation, this will lean heavily on the zeroth-order ($\Ud=0$) solutions derived in \app{appendix:zeroth-order}, which we denote using subscripts/superscripts [e.g., $\Rbo$, $\vo$, $\fo(x)$]. 

\subsection{Rare driver mutations}

\noindent In the rare driver regime, adaptation is a highly non-equilibrium process, since drivers arise and fix on very different timescales. This non-equilibrium behavior will play a crucial role in determining the effects of deleterious passengers as we will now demonstrate. If we assume that the last successful driver fixed $t$ generations ago, then the leading order deleterious corrections to $f(X,t)$ are given by \eq{eq:sweep-f} in the main text, which has mean fitness $\overline{X}(t) = - \Ud \left( 1 - e^{-\sd t} \right)$. When a new driver arises, it creates a new subpopulation $g(X)$ that sweeps through the population provided that it survives genetic drift (see \app{appendix:stochastic-model}). There are three different types of successful drivers:
\begin{enumerate}
\item The driver occurs on an $f(0)$ background and $g(\sb)$ survives drift. This is a classic ``unburdened'' sweep, since the $g(\sb)$ lineage will come to dominate the population. We denote the probability of this event by $f(0,t) \pfix(0,0,t)$.   
\item The driver occurs on an $f(0)$ background and $g(\sb)$ goes extinct, but not before it creates an additional deleterious mutation in $g(\sb-\sd)$ that survives drift. The driver mutation is successful, but it carries a deleterious passenger. We denote the probability of this event by $f(0,t) \pfix(0,1,t)$.  
\item The driver occurs on an $f(-\sd)$ background and $g(\sb-\sd)$ survives drift, so the deleterious background hitchhikes to fixation. We denote the probability of this event by $f(-\sd,t) \pfix(1,1,t)$.
\end{enumerate}
From our analysis in \app{appendix:stochastic-model}, the fixation probabilities are given by the long-time behavior of $\phi(X,\tau)$:
\begin{subequations}
\begin{align}
& \lim_{t' \to \infty} \phi(\sb,\tau=t') = \begin{cases}
\pfix(0,0,t) + \pfix(0,1,t) & \text{if $\phi(\sb-\sd,0) > 0$,} \\
\pfix(0,0,t) & \text{if $\phi(\sb-\sd,0) = 0$,}
\end{cases} \\
& \lim_{t' \to \infty} \phi(\sb-\sd,\tau=t') = \pfix(1,1,t) \, , 
\end{align} 
\end{subequations}
where $\phi(X,\tau)$ satisfies
\begin{subequations}
\begin{align}
-\frac{\partial \phi(\sb)}{\partial \tau} & = [\sb - e^{-\sd(t+t'-\tau)}] \phi(\sb) + \Ud \phi(\sb-\sd) - \frac{\phi(\sb)^2}{2} \, , \\
-\frac{\partial \phi(\sb-\sd)}{\partial \tau} & = \left[ \sb-\sd - \Ud e^{-\sd (t+t'-\tau)} \right] \phi(\sb-\sd) + \Ud \phi(\sb-2\sd) - \frac{\phi(\sb-\sd)^2}{2} \, ,
\end{align}
\end{subequations}
This system of equations is difficult to solve in general \citep{johnson:barton:2002}, but they are straightforward to solve perturbatively in the limit that $\Ud$ is small. To that end, we rewrite $\phi(x)$ in the form $\phi(X,\tau) = \phi_0(X) 1 + \frac{\Ud}{\sb} \phi_1(X)$. As described in \app{appendix:zeroth-order}, the zeroth-order stationary solutions are given by
\begin{align}
\phi_0(\sb) = \begin{cases}
0 \\
2\sb
\end{cases} \, , \quad \phi_0(\sb-\sd) = \begin{cases} 
0 \\
2(\sb-\sd)\theta(\sb-\sd) 
\end{cases}
\end{align}
and the first-order correction $\phi_1(\sb,\tau)$ satisfies the linearized equation
\begin{align}
-\partial_\tau \phi_1(\sb,\tau) = -\sb \phi_1(\sb,\tau)-2\Ud\sb e^{-\sd (t+t'-\tau)} + \begin{cases} 
2\Ud(\sb-\sd) \theta(\sb-\sd) & \text{if $\phi(\sb-\sd) > 0$,} \\
0 & \text{if $\phi(\sb-\sd) = 0$}. 
\end{cases}
\end{align} 
with the initial condition $\phi_1(\sb,0) = 0$. This equation can be solved using elementary methods, and we find that 
\begin{subequations}
\begin{align}
\pfix(0,0,t) & =  2\sb \left( 1 - \frac{\Ud e^{-\sd t}}{\sb+\sd} \right) + \mathcal{O}(\Ud^2) \, , \\
\pfix(0,1,t) & =  2 \Ud \left(1-\frac{\sd}{\sb} \right)\theta(\sb-\sd)+ \mathcal{O}(\Ud^2) \, , \\ 
\pfix(1,1,t) & = 2 (\sb-\sd) \theta(\sb-\sd) + \mathcal{O}(\Ud) \, .
\end{align}
\end{subequations}
Given these fixation probabilities, successful sweeps arise as an inhomogeneous Poisson process with rate  
\begin{align}
R(t) & = \underbrace{N\Ub[1-f_1(t)] p_\mathrm{fix}(0,0,t)}_{R_0(t)} + \underbrace{N\Ub[1-f_1(t)] p_\mathrm{fix}(0,1,t) + N\Ub f_1(t) p_\mathrm{fix}(1,1,t)}_{R_1(t)} \, ,
\end{align}
where $R_1(t)$ and $R_0(t)$ are the rates of burdened and unburdened sweeps, respectively. 
The average time between sweeps is given by
\begin{align}
\langle t \rangle & = \int_0^\infty dt \, e^{-\int_0^t dt' \, R(t') } 
 = \frac{1}{\Rbo} \left[ 1 + \frac{\Ud}{\sb} \left[ \frac{\sb^2}{\sd^2} + \left(1 - \frac{\sb^2}{\sd^2} \right) \theta(\sb-\sd) \right] \frac{\sd^2}{\sb(\Rbo+\sd)} \right] + \mathcal{O}(\Ud^2) \, ,
\end{align}
and the probability of a deleterious mutation hitchhiking on any given sweep is 
\begin{align}
p_h & = \int dt \, R_1(t) e^{-\int_0^t dt' R_0(t')+R_1(t')} \approx \frac{\Ud}{2N\Ub\sb+\sd} \left[ 1 - \left(\frac{\sd}{\sb} \right)^2 \right] \theta(\sb-\sd) + \mathcal{O}(\Ud^2) \, .
\end{align}
Thus, the beneficial and deleterious substitution rates are given by
\begin{subequations}
\begin{align}
\Rb & = \frac{1}{\langle t \rangle} \approx \begin{cases}
\Rbo \left[ 1 - \frac{\Ud \sd}{\sb^2} \frac{\sd}{\Rbo+\sd} \right] & \text{if $\sd < \sb$,} \\
\Rbo \left[ 1 - \frac{\Ud}{\sd} \right] & \text{if $\sd > \sb$,}
\end{cases} \\
\Rd & = \frac{p_h}{\langle t \rangle} \approx \begin{cases}
\Ud \left( 1 - \frac{\sd^2}{\sb^2} \right) \frac{\Rbo}{\Rbo+\sd} & \text{if $\sd < \sb$,} \\
0 & \text{if $\sd > \sb$,}
\end{cases}
\end{align}
\end{subequations}
in agreement with \eqs{eq:sweep-rd}{eq:sweep-rb} in the text. 

\subsection*{Multiple driver mutations}

\noindent The deleterious corrections in the multiple driver regime are more straightforward to calculate, since the time-dependence of $f(X,t)$ is already accounted for in the steady-state traveling wave. Substituting \eq{eq:perturbative-expansions} into \eqs{eq:f-equation}{eq:w-equation} and equating powers of $\Ud/\xc$, we obtain a set of linearized equations for $g(x)$ and $h(x)$,
\begin{subequations}
\begin{align}
\partial_x g(x) & = \eta \frac{\partial_x \wo(x)}{\wo(x)} - \frac{\wo(x) g(x)}{2 \vo} + \left( \frac{\xc}{\vo} \right) \left[ \frac{\wo(x-\sd)}{\wo(x)} - 1 \right] + \frac{\Ub}{\vo} \int ds \, \rhob(s) \left[ \frac{\wo(x+s) g(x+s)}{\wo(x)} - g(x) \right]  \, , \\
\partial_x h(x)  & = \eta \frac{\partial_x \fo(x)}{\fo(x)} + \left(\frac{\xc}{\vo} \right) \left[ \frac{\fo(x+\sd)}{\fo(x)} - 1 \right] + \frac{\Ub}{\vo} \int ds \, \rhob(s) \left[ \frac{\fo(x-s)h(x-s)}{\fo} - h(x) \right] \, ,
\end{align}
\end{subequations}
where we have defined $\eta = \etarb + \frac{\xc \sd}{\vo} \etard$. A similar expansion of the consistency condition, $\pfix(0) = \int f(x) w(x) \, dx \approx \frac{1}{N}$, yields a third relation,
\begin{align}
\label{eq:oskar-implicit-eta}
0 = \frac{\int \fo(x) \wo(x) g(x) \, dx}{\int \fo(x) \wo(x)} + \frac{\int \fo(x) [h(x) - \int \fo h] \wo(x) \, dx}{\int f_0(x) w_0(x)} \, , 
\end{align}
which uniquely determines $g(x)$, $h(x)$, and $\eta$. To carry out this calculation, it will be useful to separate $g(x)$ and $h(x)$ into parts that depend on $\eta$ and parts that depend on $\sd$:
\begin{align}
g(x) = \eta \geta(x) + \gs(x) \, , \quad h(x) = \eta \heta(x) + \hs(x) \, ,
\end{align} 
where the individual components satisfy 
\begin{subequations}
\begin{align}
\partial_x \geta(x) & = \partial_x \log \wo(x) - \frac{\wo(x) \geta(x)}{2 \vo} + \frac{\Ub}{\vo} \int ds \, \rhob(s) \left[ \frac{\wo(x+s) \geta(x+s)}{\wo(x)} - \geta(x) \right] \, , \\
\partial_x \gs(x) & = - \frac{\wo(x) \gs(x)}{2 \vo} + \frac{\xc}{\vo} \left[ \frac{\wo(x-\sd)}{\wo(x)} - 1 \right] + \frac{\Ub}{\vo} \int ds \, \rhob(s) \left[ \frac{\wo(x+s) \gs(x+s)}{\wo(x)} - \gs(x) \right] \, , \\
\partial_x \heta(x) & = \partial_x \log \fo(x) + \frac{\Ub}{\vo} \int ds \, \rhob(s) \left[ \frac{\fo(x-s)\heta(x-s)}{\fo} - \heta(x) \right] \, , \\
\partial_x \hs(x) & = - \left(\frac{\xc}{\vo}\right) \left[ \frac{\fo(x+\sd)}{\fo(x)} - 1 \right] + \frac{\Ub}{\vo} \int ds \, \rhob(s) \left[ \frac{\fo(x-s)\hs(x-s)}{\fo} - \hs(x) \right] \, ,
\end{align}
\end{subequations}
Substituting these definitions into the consistency condition in \eq{eq:oskar-implicit-eta}, we transform the implicit equation for $\eta$ into an explicit formula 
\begin{align}
\eta = - \left[ \frac{ \frac{\int f_0(x) w_0(x) \gs(x) \, dx}{\int f_0(x) w_0(x) \, dx} + \frac{\int f_0(x) [\hs(x)-\int f_0 \hs] w_0(x) \, dx}{\int f_0(x) w_0(x) \, dx}}{\frac{\int f_0(x) w_0(x) \geta(x) \, dx}{\int f_0(x) w_0(x) \, dx} + \frac{\int f_0(x) [ \heta(x) - \int f_0 \heta] w_0(x) \, dx}{\int f_0(x) w_0(x) \, dx}} \right] \label{eq:oskar-eta-consistency} \, .
\end{align}
We must then simply solve for the $g$'s and $h$'s and evaluate the required integrals. 

We begin by focusing on $\geta(x)$. Here, the solution is identical to the ``boost'' transformation discussed in \app{appendix:zeroth-order}, with $\epsilon = - \frac{\Ud}{\xc} \eta$. Thus, we can immediately conclude that
\begin{align}
\geta(x) = \begin{cases}
0 & \text{if $x > \xc$,} \\
\frac{x^2}{2 \vo} + \frac{\xc^2}{2 \vo} \left[ 1 - \frac{\sb^*}{\xc} \right] & \text{if $x < \xc$.}
\end{cases}
\end{align}
Similarly for $\heta(x)$, we find that $\heta(x) = - \frac{x^2}{2 \vo} + C$, and after renormalizing, 
\begin{align}
\heta(x) - \int \fo(x) \heta(x) = - \frac{x^2}{2\vo} + \frac{1}{2} \approx - \frac{x^2}{2\vo} \, .
\end{align}
Combining these expressions for $\geta(x)$ and $\heta(x)$ we can simplify the denominator of \eq{eq:oskar-eta-consistency}, so that
\begin{align}
\eta \approx - \left[ \frac{ \frac{\int \fo(x) \wo(x) \gs(x) \, dx}{\int \fo(x) \wo(x) \, dx} + \frac{\int \fo(x) \left[ \hs(x) - \int \fo \hs \right] \wo(x) \, dx}{\int \fo(x) \wo(x) \, dx}}{\frac{\xc^2}{2 \vo} \left[ 1 - \frac{\sb^*}{\xc} \right]} \right] \, . \label{eq:simplified-oskar-eta-consistency}
\end{align}
Now we turn our attention to $\gs(x)$ and $\hs(x)$. For the latter, we find that 
\begin{align}
\hs(x) & = \frac{x \xc}{\vo} + \frac{\xc}{\sd} e^{-\frac{x \sd}{\vo} - \frac{\sd^2}{2 \vo}} + C \, ,
\end{align}
and hence after renormalization
\begin{align}
\hs(x) - \int \fo(x) \hs(x) = \frac{x \xc}{v} - \frac{\xc}{\sd} \left[ 1 - e^{-\frac{x \sd}{\vo} - \frac{\sd^2}{2\vo}} \right] \, .
\end{align}
For $\gs(x)$, we can neglect the first-derivative term when $x > \xc$, so that
\begin{align}
\gs(x) = -\left( \frac{\xc}{x} \right) \left[ 1 - \frac{\wo(x-\sd)}{2x} \right] \, .
\end{align}
Meanwhile, for $x < \xc$ we can neglect the $\wo(x) \gs(x)$ term and obtain $\gs(x)$ by direct integration:
\begin{align}
\gs(x) & = - \frac{x \xc}{\vo} - \frac{\xc}{\sd} \theta(\sb^* - \sd) e^{-\frac{\sd \mathrm{max}\left\{x,\xc-\sb^*+\sd \right\}}{\vo}+\frac{\sd^2}{2\vo}} + C \, .
\end{align}
To solve for $C$, we return to the same integral transform that we used for the symmetry transformations in \app{appendix:zeroth-order}. For infinitesimal translations, we already saw that the $\frac{x \xc}{\vo}$ terms do not contribute to $C$ in this regime. The remaining terms yield 
\begin{align}
\begin{aligned}
C = & \left[ \sb^* \vo e^{-\frac{\xc^2}{2\vo}} \right]^{-1} \left[ - \Ub \int_{\xc-\sb^*}^{\xc} e^{-\frac{(x-\sb^*)^2}{2\vo}} \wo(x) \frac{\xc \theta(\sb^* - \sd)}{\sd} e^{-\frac{\sd \mathrm{max} \{x,\xc - \sb^* + \sd\}}{\vo} + \frac{\sd^2}{2 \vo}} \right. \\
	& \left. + \int_{\xc-\sb^*}^{\xc} e^{-\frac{x^2}{2\vo}} \wo(x)^2 \frac{\xc \theta(\sb^*-\sd)}{\sd} e^{-\frac{\sd \mathrm{max} \left\{ x, \xc-\sb^*+\sd \right\}}{\vo} + \frac{\sd^2}{2\vo}} + x_c \int e^{-\frac{(x+\sd)^2}{2\vo}} \wo(x) \right] \, ,
	\end{aligned}
\end{align}
and hence
\begin{align}
\frac{C}{\frac{\xc^2}{\vo} \frac{\Rd}{\Ud}} & = 1 - \theta(\sb^*-\sd) \left( \frac{e^{\frac{\sd^2}{v_0}}}{e^{\frac{\sb^* \sd}{\vo}} -1} \right) \left[ \frac{-1}{1-\frac{\sd}{\xc}} + \left(1-\frac{\sb^*}{\xc}\right) \left( \frac{1-e^{-\frac{\left( \sb^* - \sd \right)^2}{\vo}}}{1-\frac{\sd}{\sb^*}} +  e^{-\frac{\left(\sb^*-\sd\right)^2}{\vo}} - e^{-\frac{{\sb^*}^2-\sd \left( \sb^*-\sd\right)}{\vo}} \right) \right] \, .
\end{align}
Due to the separation of fitness scales $\vo / (\xc-\sb^*) \ll \sqrt{\vo}$ and $\vo/\sb^* \ll \sqrt{\vo}$, we can keep only the zeroth order terms in $\sd/\sqrt{\vo} \gg \sd/\sb^* \gg \sd/\xc$, which yields
\begin{align}
C = \frac{\xc^2}{\vo} \frac{\Rd}{\Ud} \left[ 1 + \frac{\sb^*}{\xc} \left( e^{\frac{\sb^* \sd}{\vo} } - 1 \right)^{-1} \right] \, .
\end{align}
Putting all of this together in our expression for $\eta$, we see that the $x \xc/\vo$ terms from the $\gs(x)$ and $\hs(x)$ integrals cancel, and we obtain  
\begin{align}
\eta & = -\frac{2 \vo}{\xc^2} \left( 1- \frac{1}{q} \right)^{-1} \left[ C - \frac{\xc}{\sd} + \frac{\xc}{\sd \sb} \int_{\xmin}^{\xc} \theta(\xc-x-\sd) e^{-\frac{x \sd}{\vo} - \frac{\sd^2}{2\vo}} \, dx \right. \nonumber \\
	& \quad \left. - \int_{\xmin}^{\xc} \theta(\xc-\xmin-\sd) e^{-\frac{\sd \mathrm{max}\{x,\xmin+\sd\}}{\vo} + \frac{\sd^2}{2\vo}} \, dx \right] \, , \\
	& = -\frac{2 \vo}{\xc^2} \left( 1- \frac{1}{q} \right)^{-1} \left[ C - \frac{\xc}{\sd} - \frac{\xc}{\sb} e^{-\frac{(\xc - \sb)\sd}{\vo} - \frac{\sd^2}{2\vo}} \right] \, .
\end{align}
The last term can be neglected since $\sd \ll \sb^*$, and we find that 
\begin{align}
\eta = 2 \left( 1 - \frac{1}{q} \right)^{-1} \left[ \frac{\vo}{\xc \sd} - \frac{\Rd}{\Ud} \left( 1 + \frac{1}{q} \frac{e^{-\frac{\sd \sb}{\vo}}}{ 1 - e^{-\frac{\sd \sb}{\vo}} } \right) \right] \, .
\end{align}

\subsection{The infinitesimal limit}

\noindent The leading-order corrections in the infinitesimal limit can be obtained in a similar manner. In this case, however, it makes more sense to partition the fitness effects into deleterious and infinitesimal components,
\begin{align}
U\rho(s) = \Ui \rhoi(s) + \Ud \rhod(-s) \, ,
\end{align}
rather than the beneficial/deleterious division in \eq{eq:dfe-partition}. Here, $\rhoi(s)$ can also contain deleterious fitness effects as long as they are sufficiently close to the infinitesimal limit ($\tc |s| \ll 1$). Recall that the mutational diffusion constant [$2 D \equiv \Ui \int s^2 \rhoi(s) \, ds$] does not depend on the sign of $s$, so these results will also apply when the fitness of the population is declining due to Muller's ratchet. Provided that $\Ud \ll \xc$, the substitution rate for the remaining deleterious mutations is given by
\begin{align}
\etard & \approx \frac{\int_{-\infty}^{\xc} f(x) w(x-\sd) \, dx}{\int_{-\infty}^{\xc} f(x) w(x) \, dx} \approx e^{-\frac{\sigma^2 \sd}{2 D}} \left( \frac{\int_{z_0}^\infty \mathrm{Ai}(z) \mathrm{Ai}\left(z+\frac{\sd}{D^{1/3}} \right)}{\int_{z_0}^\infty \mathrm{Ai} (z)^2 \, dz } \right) \approx e^{-\frac{2 \xc \sd}{\sigma^2}} \, , \label{eq:infinitesimal-etard}
\end{align}
where the last approximation captures the interesting dependence in the large-$N$ limit where $\sigma^2 \gg D^{2/3}$. 
\citet{neher:hallatschek:2013} have recently shown that the coalescent timescale is given by $\tc \sim \frac{\sigma^2}{2\xc}$ in this regime, so for fixed $\tc$, \eq{eq:infinitesimal-etard} is identical to \eq{eq:approx-etard} derived in the text. In this case, however, we see that
\begin{align}
\tc \approx \frac{\sigma^2}{2\xc} \sim \frac{\log^{1/3} \left( ND^{1/3} \right)}{D^{1/3}} \, ,
\end{align}
which increases weakly with population size. This implies that larger populations will tend to fix fewer and more weakly deleterious mutations, similar to the behavior in the single-locus case.  

To calculate the change in the infinitesimal substitution rate, we must solve for the corrections to $w(x)$ and $f(x)$ in the same manner as the previous section. In this case, however, we note that the infinitesimal substitution rate formally diverges in the infinitesimal limit ($\Ui \to \infty$), so the fractional change in $\Rb$ is simply
\begin{align}
\etarb \approx 0 + \mathcal{O}\left( \frac{\xc}{\Ui} \right) \, . 
\end{align}


\end{document}